\documentclass[iop]{emulateapj}
\usepackage{amsmath}
\usepackage{amssymb}
\usepackage{verbatim}
\usepackage{graphicx}
\usepackage{morefloats}
\usepackage{float}
\usepackage{lipsum}
\usepackage{subfigure}
\usepackage{longtable}
\usepackage{lipsum} 
\usepackage{rotating}
\usepackage{url} 
\usepackage{natbib}

\newcommand{\FeII}{Fe II $\lambda$5169~} 
\newcommand{\FeIIn}{Fe II $\lambda$5169} 
\newcommand{\kms}{km s$^{-1}$}

\begin{document}

\title{Analyzing the largest spectroscopic dataset of hydrogen-poor super-luminous supernovae}

\author{Yu-Qian Liu\altaffilmark{1}, Maryam Modjaz\altaffilmark{1}, Federica B. Bianco\altaffilmark{1, 2}}
\altaffiltext{1}{Center for Cosmology and Particle Physics, New York University, 4 Washington Place, New York, NY 10003, USA; YL1260@nyu.edu; mmodjaz@nyu.edu}
\altaffiltext{2}{Center for Urban Science and Progress, New York University,
1 MetroTech Center, Brooklyn, NY 11201, USA}

\begin{abstract}
Super-luminous supernovae (SLSNe) are tremendously luminous explosions whose power sources and progenitors are highly debated. Broad-lined SNe Ic (SNe Ic-bl) are the only type of SNe that are connected with long-duration gamma ray bursts (GRBs). Studying the spectral similarity and difference between the populations of hydrogen-poor SLSNe (SLSNe Ic) and of hydrogen-poor stripped-envelope core-collapse SNe, in particular SNe Ic and SNe Ic-bl, can provide crucial observations to test predictions of theories based on various power source models and progenitor models. In this paper, we collected all of the published optical spectra of 32 SLSNe Ic, 21 SNe Ic-bl, as well as 17 SNe Ic, quantified their spectral features, constructed average spectra, and compared them in a systematic way using new tools we have developed. We find that SLSNe Ic and SNe Ic-bl, including those connected with GRBs, have comparable widths for their spectral features and average absorption velocities at all phases. Thus, our findings strengthen the connection between SLSNe Ic and GRBs. In particular, SLSNe Ic have average \FeII absorption velocities of $-15,000 +/- 2,600$ \kms~at 10 days after peak, which are higher than those of SNe Ic by $\sim7,000$ \kms~on average. SLSNe Ic also have significantly broader \FeII lines than SNe Ic. Moreover, we find that such high absorption and width velocities of SLSNe Ic may be hard to explain by the interaction model, and none of 13 SLSNe Ic with measured absorption velocities spanning over 10 days has a convincing flat velocity-evolution, which is inconsistent with the magnetar model in one dimension. Lastly, we compare SN 2011kl, the first SN connected with an ultra-long GRB, with the mean spectrum of SLSNe Ic and of SNe Ic-bl.    

\end{abstract}

\section{Introduction}
\label{intro}

Over the last 10 years, the systematic discovery of Super-luminous Supernovae (SLSNe) has captured the attention of the supernova community, causing heated debates about their powering source and their progenitors. SLSNe are defined as SNe that have peak magnitudes more luminous than around $-21$ magnitude, making them at least $\sim100$ times more luminous than normal stripped-envelope core-collapse SNe (stripped SNe) and at least $\sim10$ times more luminous than SNe Ia \citep{gal-yam12, gal-yam16}. Based on their spectra, SLSNe are classified into Type I SLSNe (SLSNe I) that do not show hydrogen lines in their spectra near peak, and Type II SLSNe (SLSNe II) that do show hydrogen lines \citep{quimby11, gal-yam12}. 

Currently, three distinct mechanisms have been suggested for powering the brilliance of SLSNe I: (1) pair-instability SNe \citep[e.g.,][]{Rakavy67, gal-yam09}, (2) engine-driven, either via a magnetar \citep{kasen10, woosley10, inserra13, nicholl13, Metzger15, chen16, Suzuki16, Gilkis16, Soker16} or via accretion onto a black hole \citep{Dexter13}, and (3) interaction of the supernova ejecta with a circumstellar material that is free of hydrogen and helium and that had been ejected before the explosion \citep[e.g.,][]{Chevalier11, Ginzburg12, Chatzopoulos13, ofek14}, possibly due to pulsations in the stages immediately before the explosion, as in the pulsational pair instability SN (PPISN; \citealt{Woosley07, Waldman08, Woosley16}). Recently, hybrid models have invoked several of these mechanisms for the same object to explain the peculiar light curves (e.g., PTF12dam by \citealt{Tolstov16}, iPTF13ehe by \citealt{wang16}, ASASSN-15lh by \citealt{Chatzopoulos16}). Based on current observations of SLSNe, the PISN origin has been disfavored for most of the SLSNe I \citep[except for SN 2007bi;][]{gal-yam09} for a number of reasons. Many PISN models provide overly broad light curves and spectra that are too red compared to observations \citep[e.g.,][]{dessart12_slsn, dessart13}.

SLSNe I show spectra devoid of hydrogen (H) and helium (He) lines, \footnote{Except for SN 2012il that appears to show He I $\lambda$10830 in emission, but none of the optical He I lines \citep{inserra13} and iPTF13ehe that showed broad H$\alpha$ emission only in nebular spectra \citep{yan15}.} which is also the case for SNe Ic and SNe Ic-bl. Thus, we denote SLSNe I as SLSNe Ic in this work, following the same name convention as in other works (e.g., \citealt{inserra13}). The similarity in the absence of absorption lines of H and He motivates us to compare spectra of SLSNe Ic, SNe Ic, and SNe Ic-bl. A basic question is whether they are truly distinct populations, or whether there is a smooth transition between them. Given the large number of SLSN Ic spectra, as well as recent systematic explorations of other members of the SN Ic family (i.e, SNe Ic and Ic-bl) by us \citep{liu16, modjaz15}, the time is ripe to conduct a similar systematic spectroscopic population study for SLSNe Ic. In Section \ref{data}, we summarize SN samples used in this study and discuss the potential caveats of our samples. In Section \ref{method}, we discuss our spectral analysis methods and report our measurements of absorption velocities as well as the construction of average spectra. In Section \ref{connection}, we conduct spectral comparisons between SLSNe Ic, SNe Ic-bl and SNe Ic. In Section \ref{comp_model}, we compare measured velocities in this work with predicted velocities in the interaction model and the one-dimensional magnetar model. In Section \ref{weird_spec}, we compare several special SLSNe Ic to average spectra to see if these SLSNe Ic are typical SLSNe Ic. Finally, we summarize our conclusions in Section \ref{summary}. 

\section{SN spectral samples}
\label{data}

We list our SLSN Ic sample in Table \ref{table}. Our samples of SNe Ic and of SNe Ic-bl, roughly half of which were connected with GRBs, are the same as in \citet{modjaz15}. Since we want to analyze SLSN Ic spectra as a function of phase in a statistical way, we have collected the spectra of all available SLSNe Ic that have a date of maximum light published before April of 2016. These SLSNe Ic were mainly discovered and observed by the All-Sky Automated Survey for Supernovae (ASAS-SN), the Catalina Real-Time Transient Survey (CRTS), the Dark Energy Survey (DES), the Hubble Space Telescope Cluster Supernova Survey, the Pan-STARRS1 Medium Deep Survey (PS1), the Public ESO Spectroscopic Survey of Transient Objects (PESSTO), the Intermediate Palomar Transient Factory (iPTF) as well as the Palomar Transient Factory (PTF), and the Supernova Legacy Survey (SNLS). 

We have included two special SLSNe Ic, namely SN 2011kl, which constitutes the first detection of a supernova explosion associated with an ultra-long duration gamma ray burst (ULGRB), and ASASSN-15lh (also known as - AKA - SN 2015L), which is claimed to be the most luminous SN ever discovered, although whether it is a SN or not is still debated. We did not include transients in the SN-SLSNe gap \citep{Arcavi16}, which include PTF10iam, SNLS04D4ec, SNLS05D2bk, and SNLS06D1hc, since the first object is a weird type II SN, the following two objects have no spectra classifications, and the last object only has a featureless spectrum obtained three weeks after the explosion. We did not include PS1-10afx, which was first classified as a SLSN Ic \citep{chornock13}, but turned out to be a lensed SN Ia \citep{Quimby13}.

The question of whether all SLSNe Ic are aspherical explosions is an important one, and includes whether their geometry would introduce a selection effect for sample comparison purposes. Observationally, there are only few SLSNe Ic for which the geometry of the explosion has been directly constrained. Out of two objects that have polarimetric observations, only one, SN2015bn, appears to show strong polarization \citep{Inserra16, Leloudas17}. Late time spectra and observations of SN remnants (e.g., \citealt{Omand2017}) can also be used to probe the geometry of the explosions. For example, \citet{Inserra17} suggested multiple emission regions for the same emission line in the late time spectra of LSQ14an. However, this could be due to asphericity of the ejecta or due to interactions, and thus, it is not clear whether LSQ14an was a large-scale aspherical explosion. Thus, we conclude that while certainly more data are needed to address the question of asphericity in SLSNe Ic, we have good reasons to believe based on current data that selection effects will probably not be significant for our sample.

We note that although every SLSN Ic in our sample has a date of maximum light, the reference band is not consistent. The potential effects of different peak epochs in different filters will be discussed in subsection \ref{diff_filter}. Another caveat is that since some SLSNe Ic are at high redshifts, their spectra taken at optical wavelengths may not cover the full optical range when the observed spectra are converted to rest wavelength. We will discuss the potential effects of different redshifts for different SN samples in subsection \ref{diff_redshift}.

In summary, we have 207 optical spectra of 17 SNe Ic, 200 optical spectra of 21 SNe Ic-bl, and 178 optical spectra of 32 SLSNe Ic. Here, we report the mean and standard deviation redshifts, as well as median redshifts, for each of the SN subtypes in our sample: $z_\mathrm{Ic, mean}=0.02\pm0.03$, $z_\mathrm{Ic, median}=0.01$, $z_\mathrm{Ic-bl, mean}=0.09\pm0.13$, $z_\mathrm{Ic-bl, median} = 0.03$, $z_\mathrm{SL, mean}=0.47\pm0.39$, and $z_\mathrm{SL, median} = 0.33$.   

\subsection{Potential Impact of Different Date of Maxima in Different Rest Frame Filters}
\label{diff_filter}
Table \ref{table} lists the dates of maximum light, which were provided by the published papers in different filters for the different SLSNe Ic. Most of the rest-frame filters are in $u$, $B$, $g$, $r$, or $R$-bands. Since there are no established relationships for the peak epochs in different filters of SLSNe Ic, we could not convert these peak epochs to the same filter. Since there is no common filter in which light curves of all SLSNe Ic in our sample are available in the literature, we could not compute peak epochs in the same filter. Based on the relationship between peak epoch in $U$, $B$, $V$, and $R/r'$-bands for stripped-envelope core-collapse SNe from \citet{bianco14}, we roughly estimate that compared to peak epoch in $V$-band (JD$_{\mathrm{vmax}}$), the peak epochs used in our analyses will be off by $-3$ days (peak happened before the JD$_{\mathrm{vmax}}$) and $2$ days the most (peak happened after the JD$_{\mathrm{vmax}}$). This should not impact our comparisons of average spectra (see Section \ref{AveSpec} and sections that used average spectra of SLSNe Ic), since we use a bin size of 5 days for each average spectrum. Neither should this have much influence on our analysis of the bulk velocity evolution (see Section \ref{velFeII}), since we calculate the moving weighted average velocities smoothed with a Gaussian kernel with a standard deviation of 5 days when studying the velocity evolution of the different SN populations. 

\subsection{Potential Impact of Different Redshift Distributions for Different SN subtypes}
\label{diff_redshift}
The different SN samples in this work are at vastly different redshifts with
means of $z_\mathrm{Ic, mean}=0.02$, $z_\mathrm{Ic-bl, mean}= 0.09$, $z_\mathrm{SL}=0.47$ and medians of $z_\mathrm{Ic, median}=0.01$, $z_\mathrm{Ic-bl, median} = 0.03$, $z_\mathrm{SL, median} = 0.33$, thus a valid question is whether the different redshift coverage has any impact on our results. At the broadest level, this should not influence our results too much, since all SN spectra are de-redshifted into the rest frame, and all spectral comparisons are made for the same spectral ranges (e.g., the expected wavelength range for the \FeII line) that are observed in almost all SN spectra in our sample. However, five SLSNe Ic in our sample that have $z>0.9$ do not cover the \FeII region when their spectra are de-redshifted into the rest frame, since the \FeII line will be at infrared wavelengths in the observed frame. Moreover, the higher redshifts of the SLSNe Ic sample does somewhat impact our comparison of average spectra (see Section \ref{AveSpec}), since those of SLSNe Ic have a larger UV and blue coverage in the rest-frame ($\sim3$000 \AA~- 7000 \AA) than those of SNe Ic ($\sim4$000 \AA~- 8000 \AA). Nevertheless, for the important blend of O II at $\sim4$300 \AA~seen in spectra of SLSN I, we have spectra of SNe Ic and SNe Ic-bl at the same rest wavelengths  (see Figure \ref{OII_mont}). 

The next question is whether evolutionary effects for their progenitors is expected to be significant within such a redshift range. While the metallicity content of the universe does not change significantly between the redshift ranges for the SN Ic and the SN Ic-bl samples \citep[both samples are within the extent of the star-forming galaxies of SDSS;][]{tremonti04}, there is a slight trend towards lower metallicity for galaxies of the same mass at the median redshift of $z=0.3$ for SLSNe Ic \citep[e.g.,][]{Zahid11, Maier15}, compared to local star-forming galaxies. Indeed, SLSNe Ic host galaxies seem to be of low metallicity \citep{chen13, lunnan13, Lunnan14, chen2016, perley16, Schulze16}, even lower than those of SNe-GRBs \citep{Leloudas15}, which in turn seem to be lower than those of SNe Ic and SNe Ic-bl without GRBs \citep{modjaz08_Z, modjaz11, sanders12, kelly12, modjaz12_proc}. However, low metallicity should not affect the velocities of the \FeII line per se. Even the \FeII line strengths should be unchanged for our sample of SLSNe Ic, since line strength changes only become important when the metallicity is by factors of 10 to 100 lower \citep{Sauer06}, which is a much larger factor than observed based on host galaxy data.

\begin{deluxetable*}{lcccl}
\tablecolumns{3}
\singlespace
\tablecaption{Spectral sample of SLSNe Ic}
\tablehead{
\colhead{SN Name\tablenotemark{a} } &
\colhead{Redshift $z$} &
\colhead{Phases\tablenotemark{b} of Spectra}  &
\colhead{LC Band\tablenotemark{c}}&
\colhead{References\tablenotemark{d}}
}
\startdata
ASASSN-15lh\tablenotemark{$\star$} & 0.2326 &13, 15, 20, 26, 36,
39 & V (B)&\citet{dong16}  \\  [0.5ex]
~~(or SN 2015L)\\[1ex]
DES13S2cmm & 0.663 &35 & r (u)&\citet{Papadopoulos15}\\ [0.5ex]
~~(or SN 2013hy) &&\\[1ex]
DES14X3taz & 0.608 &$ -20,
-13$ & g (u)  &\citet{smith16}\\ [1ex]
LSQ12dlf &0.25 &6, 8, 8, 15, 20, 33,
43 & V (B)&\citet{nicholl14}\\ [1ex]
LSQ14bdq &0.35 &$-19$ & r (g) &\citet{nicholl15_lsq14bdq}\\ [1ex]
LSQ14mo & 0.256&$ -7, -2$, 9, 15, 22 &r (g) &\citet{chen2016}\\ [1ex]
PS1-10awh & 0.908&$-18$, 6,
13 & mix of filters\tablenotemark{$\dagger$}&\citet{chomiuk11} \\ [1ex]
PS1-10bzj & 0.650 &7, 14,
16 & bolometric &\citet{lunnan13}\\ [1ex]
PS1-10ky & 0.956&$ -2$, 1, 15,
26 &  mix of filters\tablenotemark{$\dagger$}&\citet{chomiuk11}\\ [1ex]
PS1-11ap & 0.524&$ -20, -1$, 10, 38, 78,
90+(1) &  r (g)&\citet{McCrum14}\\ [1ex]
PTF09atu & 0.501&$-20$ & (u)&\citet{quimby11}\\ [1ex]
PTF09cnd & 0.258&$-18$,
90+(1) & (u)&\citet{quimby11}\\ [1ex]
PTF09cwl& 0.349&$-2$ & (u)&\citet{quimby11}\\ [0.5ex]
~~(or SN 2009jh) &&\\[1ex]
PTF10cwr  & 0.231&$ -8, -5, -4$, 4, 10, 21, 29,
57 & (u)&\citet{quimby11} \\ [0.5ex]
~~(or SN 2010gx) &&\\[1ex]
PTF10hgi  & 0.0985& 33, 68,
81 & g (g)&\citet{inserra13}\\ [0.5ex]
~~(or SN 2010md)&&\\[1ex]
PTF11rks & 0.193& 2, 9, 17,
48 & g (g)&\citet{inserra13}\\ [1ex]
PTF12dam & 0.1078&$ -16, -15, -14, -14, -8, -1,$ 10, 14, 14, & bolometric &\citet{nicholl13}\\ [0.5ex]
&&18, 34, 55, 65, 65, 90+(5)& \\[1ex]
SCP06F6 & 1.189&$ -11$, 0,
5 & z (g)&\citet{Barbary09}\\ [1ex]
SNLS06D4eu & 1.588&$-15$ & bolometric&\citet{howell13} \\ [1ex]
SNLS07D2bv & 1.5&$-4$ &i (u)&\citet{howell13} \\ [1ex]
SSS120810-23 & 0.17 & 7, 8, 13, 32, 39, 
57 & R (R)&\citet{nicholl14}\\ [1ex]
iPTF13ajg & 0.74&$ -9, -8, -4$, 8, 9, 25, 27, 46,
80 &  R (B)&\citet{Vreeswijk14}\\ [1ex]
iPTF13ehe & 0.33 &$ -9,
-5$, 14, 90+(3) &  r (g)&\citet{yan15} \\ [1ex]
SN 2005ap & 0.2832&$ -2$, 5,
5 &  unfiltered&\citet{quimby07_05ap}\\ [1ex]
SN 2006oz & 0.376&$-5$ & bolometric&\citet{leloudas12}\\ [1ex]
SN 2007bi & 0.1279& 48, 54, 55, 90+(4) & R (R)&\citet{gal-yam09}\\ [0.5ex]
& & & &\citet{young10}\\ [1ex]
SN 2011ke& 0.143& 9, 15, 24, 31, 38,
45 & g (g)&\citet{inserra13}\\ [0.5ex]
~~(or PTF11dij, PS1-11xk)&&\\ [1ex]
SN 2011kf& 0.245& 25,
51 & g (u)&\citet{inserra13}\\ [1ex]
SN 2011kl/GRB111209A\tablenotemark{$\star\star$}& 0.677&$-1$ &  bolometric&\citet{Greiner15}\\ [1ex]
SN 2012il & 0.175& 13, 41,
41 & g (g)&\citet{inserra13} \\ [0.5ex]
~~(or PS1-12fo)&&\\[1ex]
SN 2013dg & 0.26& 3, 5, 15, 21, 35,
46 & r (g)&\citet{nicholl14}\\ [1ex]
SN 2015bn & 0.1136&$ -28, -27, -22, -21, -18, -9, -8, -2$, 3, 6, &  r (r)&\citet{nicholl16}\\ [0.5ex]
~~(or PS15ae)& &19, 30, 31, 44, 50, 57, 71, 83, 89, 90+(3)& 
\enddata
\tablenotetext{a}{SN names are from the listed references in the last column. Other names for the same SN are indicated in parentheses. }
\tablenotetext{b}{Phases are in the rest-frame with respect to maximum light and rounded to the nearest whole day. The number in parentheses is the number of spectra with phases larger than 90 days after the date of maximum light, which we include for completeness, but do not analyze here. The references for the date of maximum light are the same as references in the last column.}
\tablenotetext{c}{Light curve (LC) filters are used to determine dates of maximum light in references in the last column. The LC filters without parentheses are in the observed-frame, while those in parentheses are in the rest-frame. The latter is either converted to rest-frame by us or taken from the references in the last column.}
\tablenotetext{d}{References are only for the SN spectra, not for the SN discovery.}
\tablenotetext{$\star$}{Whether this is a SN or tidal disruption event is still debated.}
\tablenotetext{$\star\star$}{This is the only SLSN that was discovered in connection with an ultra-long GRB.}
\tablenotetext{$\dagger$}{The authors set a date of maximum light based on LCs on $g$, $r$, $i$, $z$, $y$-filters. Thus, the date of maximum light is a rough estimate, not a true measurement.}
\label{table}
\end{deluxetable*}

\section{Spectral Analysis Methods}
\label{method}

\subsection{Absorption Velocity of \FeII}
\label{vel_sum}

Absorption velocities can provide clues about the dynamics of the explosion. In particular, the absorption velocity of \FeII has been suggested to trace photospheric velocity by \citet{branch02}.\footnote{See \citet{dessart15, dessart16} and \citet{modjaz15} for discussions on what the definition of ``photospheric" velocity is and which line to use. While there is some discussion about which line best traces the ejecta velocity, in any case, a systematic comparison of the same way of measuring the velocity from the same line for all SNe and SN subtypes is needed and yields relative values that are still powerful to test models. For more caveats, see Section \ref{caveat}.} In turn, the measurement of the photospheric velocity is needed to estimate explosion parameters, such as ejecta masses, from light curves and spectra \citep{drout11, cano13, lyman14, taddia15, nicholl15}. Moreover, the ejecta velocity as traced by \FeII can be used to check the prediction of the one-dimensional magnetar model since the model predicts a flat evolution of velocity over time \citep{kasen10, mazzali16}.

\subsubsection{Velocity from Template Fitting Method}
\label{vel}
Since the Fe II features in SLSNe Ic are highly blended, as in SNe Ic-bl, we measured \FeII velocities via the template fitting method (Appendix A of \citealt{modjaz15}), originally developed for spectra of SNe Ic-bl. First, we identify the spectral region of \FeII in SLSNe Ic in our sample based on identifications and SYNOW fits given for the same SLSNe Ic in the literature, as well as spectral modeling for some SLSNe Ic in \citet{dessart12_slsn}. Then we measure \FeII velocities in SLSNe Ic via the template fitting method described in Appendix A of \citet{modjaz15}. This method is a data-driven method and the template is the average spectrum of SNe Ic. In our case, we would like to compare the velocities of the non-blended \FeII in SNe Ic with the velocities of the blended \FeII in SNe Ic-bl and SLSN I. We obtain the \FeII velocities in SLSNe Ic by matching a broadened and blue-shifted SN Ic template to a SLSN Ic spectrum at similar phases, which is performed in a Markov-Chain-Monte-Carlo (MCMC) framework. As discussed in the Appendix A of \citet{modjaz15}, the broadness and blue-shift of the line are quantified by the convolution velocity of the Gaussian kernel with which the SN Ic template is convolved and the amount of additional blueshift, respectively. The corresponding error bars are based on marginalized distribution of parameter values from MCMC samplers. An example of our velocity measurements for SLSNe Ic is shown in Table \ref{vels_table} for guidance. A full version of this table is available in a machine-readable form in the online journal.

We could not measure \FeII velocities in all 178 spectra of 32 SLSNe Ic in our sample. We only focus on spectra taken during the photospheric phase, which include 155 spectra, using $t_\mathrm{max}=70$ days as the threshold. Some SLSNe Ic have multiple spectra at the same phase. In these cases, we included the spectrum that has the largest wavelength coverage or the highest signal to noise ratio. As a result, 122 out of 155 spectra are at unique phases. Out of the 122 spectra of 32 SLSNe Ic, we could measure \FeII velocities in 72 spectra of 21 SLSNe Ic, since 10 spectra are too noisy to identify the \FeII feature, 16 spectra do not cover the \FeII region, and 23 spectra have no significant \FeII absorption. We could not use the template fitting method to measure the \FeII velocity in SN 2007bi at phase 48 either, since the width of its \FeII profile is narrower than that in the SNe Ic average spectrum, although the shape of the former is similar to the latter. If we use the identification method, which is used to measure the \FeII velocity in SNe Ic \citep{liu16}, the velocity we obtain for SN 2007bi at phase 48 is at the high end of all SLSN Ic velocities at similar phases. However, for consistency, we did not include the \FeII velocity for SN 2007bi from identification method in our analyses. 

We note that the line feature at the expected position of \FeIIn, which is around 5000 \AA, can be contaminated by Fe III lines at early phases if the temperature is sufficiently high. For example, through detailed comparison with identifications and SYNOW fits given for literature SLSNe Ic, \citet{nicholl13} and \citet{nicholl16} identified Fe III lines around 5000 \AA ~shortly after peak in PTF12dam, PS1-11ap, SN 2010gx, and SN 2015bn. While in Table 2, we report all \FeII velocities in SLSNe Ic, assuming that the feature is not contaminated, we are cautious with those measured at $t_\mathrm{max}<10$ days. Thus, we do not report weighted average \FeII velocities in SLSNe Ic for comparison with other SN types for $t_\mathrm{max}<10$ days. The fact that for some SLSNe Ic, those velocities measured at $t_\mathrm{max}<10$ days appear to be increasing, instead of decreasing, indicates that there is certainly some contamination at early times. 

\begin{deluxetable}{c c c c}
\tabletypesize{\scriptsize}
\tablecaption{Measured absorption velocities\label{vels_table}}
\tablehead{
\colhead{Phase with respect to maximum light} &
\colhead{\FeII} & \\
\colhead{(days)} &
\colhead{(km s$^{-1}$)} &
}
\startdata
\multicolumn{2}{c}{\bf LSQ12dlf} \\
$8$ &    $-18500^{+1300}_{-1200}$  \\ [1.5ex]
$15$ &   $-16900^{+2600}_{-2600}$  \\[1.5ex]
20 & $-13100^{+1300}_{-1300}$\\[1.5ex]
33 & $-10700^{+1400}_{-1400}$\\[1.5ex]
43 & $-5000^{+800}_{-800}$
\enddata
\tablecomments{This table is available in its entirety in a machine-readable form in the online journal. A portion is shown here for guidance regarding its form and content.}

\end{deluxetable}


\subsubsection{Velocity Comparisons with Literature Measurements}
\label{velComp}
\citet{nicholl15} have measured \FeII velocities for 8 well-observed SLSNe Ic. They separately fit three Gaussians to the \FeII region with the same initial parameters for the Gaussian, but with different wavelength ranges over which the fit is performed. Then, the mean of the three measurements is taken as the velocity, and the standard deviation as the errors (Matt Nicholl, private communication 2016). We have compared our measurements for the same 8 SLSNe Ic in this work with those in \citet{nicholl15}. Although most of their measurements are consistent with ours within the error bars, there is a systematic difference. In general, before $t_\mathrm{max}\simeq25$ days, our velocities are higher than theirs, while after $t_\mathrm{max}\simeq25$ days, our velocities are lower than theirs. The systematic difference is largely due to the different methods used in \citet{nicholl15} and this work. For both methods, the assumed profiles cannot properly describe every spectrum. However, since \FeII is usually blended with other two Fe II lines in SNe with high absorption velocities, as in SNe Ic-bl and SLSNe Ic, treating the \FeII region as only due to \FeII is not appropriate. In our template fitting method, as described in detail in the Appendix A of \citet{modjaz15}, we have taken the blending effect into consideration. As we will show in Section \ref{FlatVel}, this systematic difference in \FeII velocity measurements between \citet{nicholl15} and this work will affect our conclusions about whether the velocity evolution of SLSNe Ic is flat, which is a prediction of the one-dimensional magnetar model. In addition, we analyze additional spectra from the literature that were not shown in \citet{nicholl15}, which are at the extremes of the time coverage (i.e., either very early or very late). Velocity measurements of \FeII in these spectra will also affect the trends in velocity evolution of SLSNe Ic, since they provide a larger baseline in time. 

\begin{figure*}[t]
\subfigure{%
\includegraphics[scale=0.5,angle=0]{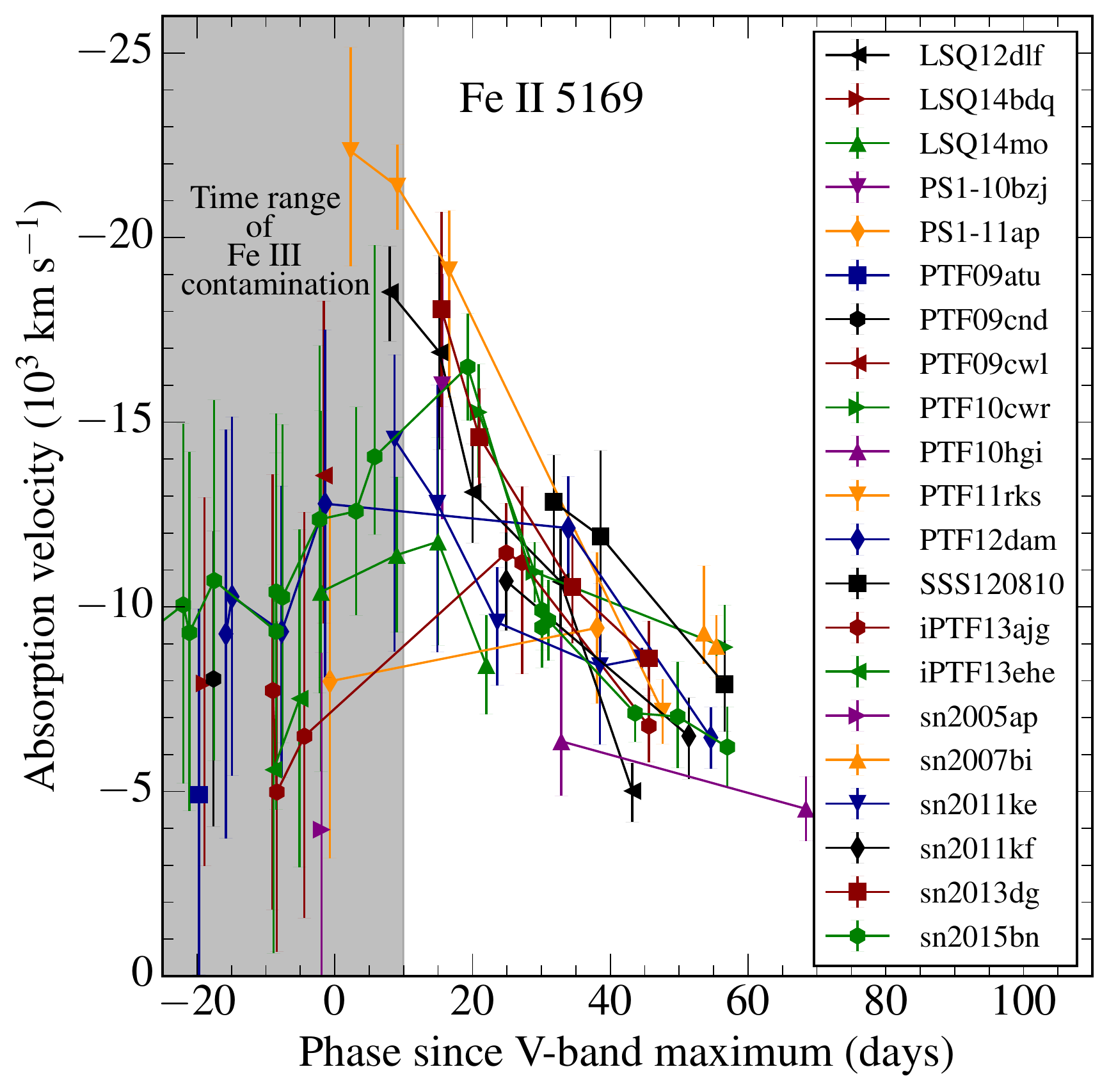}
}
\quad
\subfigure{%
\includegraphics[scale=0.5,angle=0]{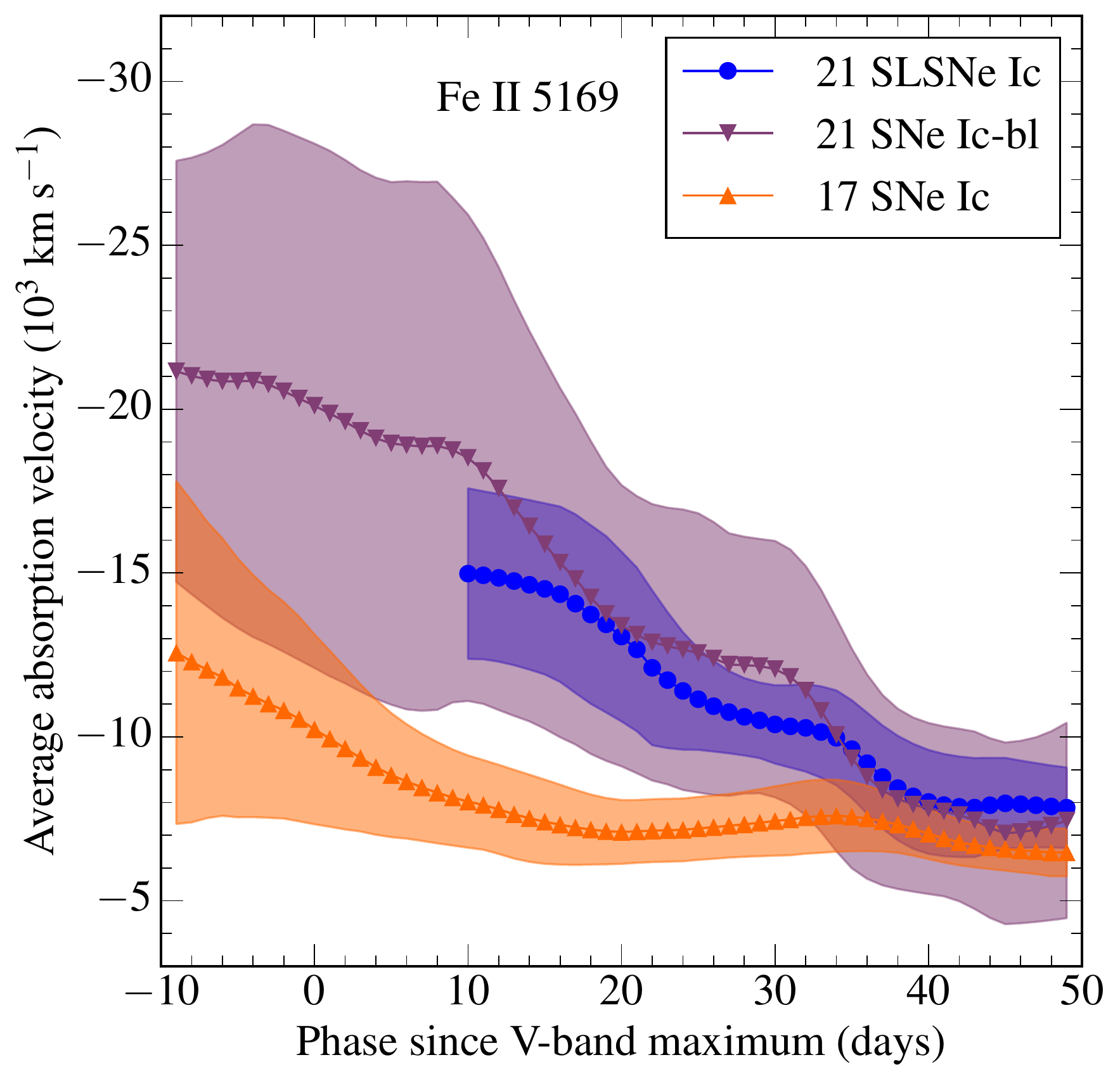}
}
\caption{\textit{Left}: Measured \FeII absorption velocities for individual SLSNe Ic. Data of the same SN are connected by a line. Data points in the shaded region are likely to be contaminated by Fe III. Note that we do not show values for DES14X3taz, sn2006oz, and ASASSN-15lh, since no obvious \FeII feature is detected in their spectra. A few SLSNe Ic have spectra that are too noisy, e.g., SN 2011kl, SN 2012il, and DES13S2cmm, or do not have spectra that cover the \FeII region, e.g., SCP06F6, SNLS06D4eu, SNLS07D2bv, PS1-10awh, and PS1-10ky, thus, we could not obtain measurements for them either. \textit{Right}: Moving weighted average velocities smoothed with a Gaussian kernel with a standard deviation of 5 days. The weights are taken to be the inverse of the square of the errors of individual measurements. Resepctively, 21 SLSNe Ic, 21 SNe Ic-bl and 17 SNe Ic are included. We only compute the average velocity for SLSNe Ic at $t_{\mathrm{Vmax}}>=10$ days, due to the possible Fe III contamination at earlier times. The error bands on the mean values represent the weighted standard deviation of the data points. 
}
\label{plot_vel}
\end{figure*}

\subsection{Constructing Mean Spectra from SN Spectra}
\label{AveSpec}

Mean spectra and the standard deviation contours can characterize the spectral properties of each SN subtype as a population. They can also be used to check whether spectra of a newly discovered transient are within the statistical diversity of the mean spectra of some SN types at certain phases. In order to compare the average spectra of various SN subtypes, as we will show in Section \ref{AveSpecSL}, we have used continuum-removed average spectra of SN Ic and of SN Ic-bl from \citet{liu16} and \citet{modjaz15}, and constructed average spectra of SLSNe Ic in the same way. In order to compare two relatively special SLSNe Ic in our sample to average spectra of various SN subtypes, as we will show in Section \ref{weird_spec}, we have also constructed average spectra where the continuum is included for SN Ic-bl, SN-GRB, and SLSN Ic at needed phases. We have done our best to correct for Milky Way extinction in SLSN Ic spectra using E(B-V) values from the literature \citep{schlegel98, Schlafly11}. We did not correct for host extinction, since SLSNe Ic are generally in metal-poor galaxies, and thus the host extinction is small in most cases \citep{Lunnan14, Leloudas15}. All of the average spectra of SLSNe Ic that we have constructed are available on our group GitHub repository. \footnote{https://github.com/nyusngroup}

Although the temperature in the photosphere of SLSNe Ic affects the presence and strength of spectral lines, we did not group SLSNe Ic by temperature when we construct their average spectra, since at the date of maximum light, SLSNe Ic generally have a blackbody temperature ($T_\mathrm{BB}$) between 10,000 K and 16,000 K, except for iPTF13ehe \citep[$\sim7$,000 K;][]{yan15}, PS1-10awh \citep[$\sim20$,000 K;][]{chomiuk11}, and ASASSN-15lh \citep[$\sim22$,000 K;][]{dong16}.


\section{Connections between SLSNe Ic, SNe Ic-bl, and SNe Ic}
\label{connection}

SLSNe Ic, SNe Ic-bl and SNe Ic are marked by spectra with no strong hydrogen absorption lines, as well as no strong helium lines \citep[though there is extensive discussion about potential Helium lines in SLSNe Ic in][]{mazzali16}. Studying how these SN subtypes are connected is a very important step in answering questions about their power sources and their progenitors \citep{Pastorello10, gal-yam12}. Although no firm physical connection has been established, \citet{Pastorello10} found that SLSN Ic PTF10cwr (AKA 2010gx) at $t_\mathrm{max}\simeq21$ days is similar to the spectrum of SN Ic-bl 2003jd at $t_\mathrm{max}\simeq2$ days and the spectrum of SN Ic at $t_\mathrm{max}\simeq-6$ days. This connection motivated us to explore any population similarity by quantifying common spectral features and generating average spectra as a function of SN subtypes in the SN Ic family, and to compare them in a systematic way. 

From line identifications and spectral modeling in various works, we know that SLSNe Ic, SNe Ic-bl and SNe Ic all show Fe II lines, but the Fe II lines can be usually identified in SLSNe Ic only after the date of maximum light, while for SNe Ic and SNe Ic-bl, they can be identified in spectra obtained starting before maximum light. As we will show in Sections \ref{velFeII} and \ref{AveSpecSL}, SLSNe Ic and SNe Ic-bl have similar spectral features and absorption velocities, which are broader and higher than those in SNe Ic. This indicates that SLSNe Ic and SNe Ic-bl may have similar explosion engines. 

Spectra of SLSNe Ic show very blue continua \citep[e.g., ][]{quimby07_05ap}. Thus, if the continuum is a black-body, the derived temperature is high enough to produce O II lines. In fact, early spectra of almost all SLSNe Ic show a narrow `w' feature near 4300 \AA, which is identified as O II possibly due to high temperatures \citep{quimby07_05ap, quimby11, gal-yam12, nicholl15, mazzali16} and non-thermal excitation \citep{mazzali16}. As we will show in Section \ref{OII}, this O II feature is not seen in SNe Ic-bl and SNe Ic. Actually, the narrow O II feature in SLSNe Ic is surprising, since we find that the \FeII feature and other features later on are broad lines. We speculate that velocities of the `w' feature do not reflect the global dynamics, but may be constrained to a small region in the outer layers where the conditions (e.g. ionization and/or temperature) are conducive.

\begin{figure*}[t]
\subfigure{%
\includegraphics[width=0.5\textwidth,scale=0.5,angle=0]{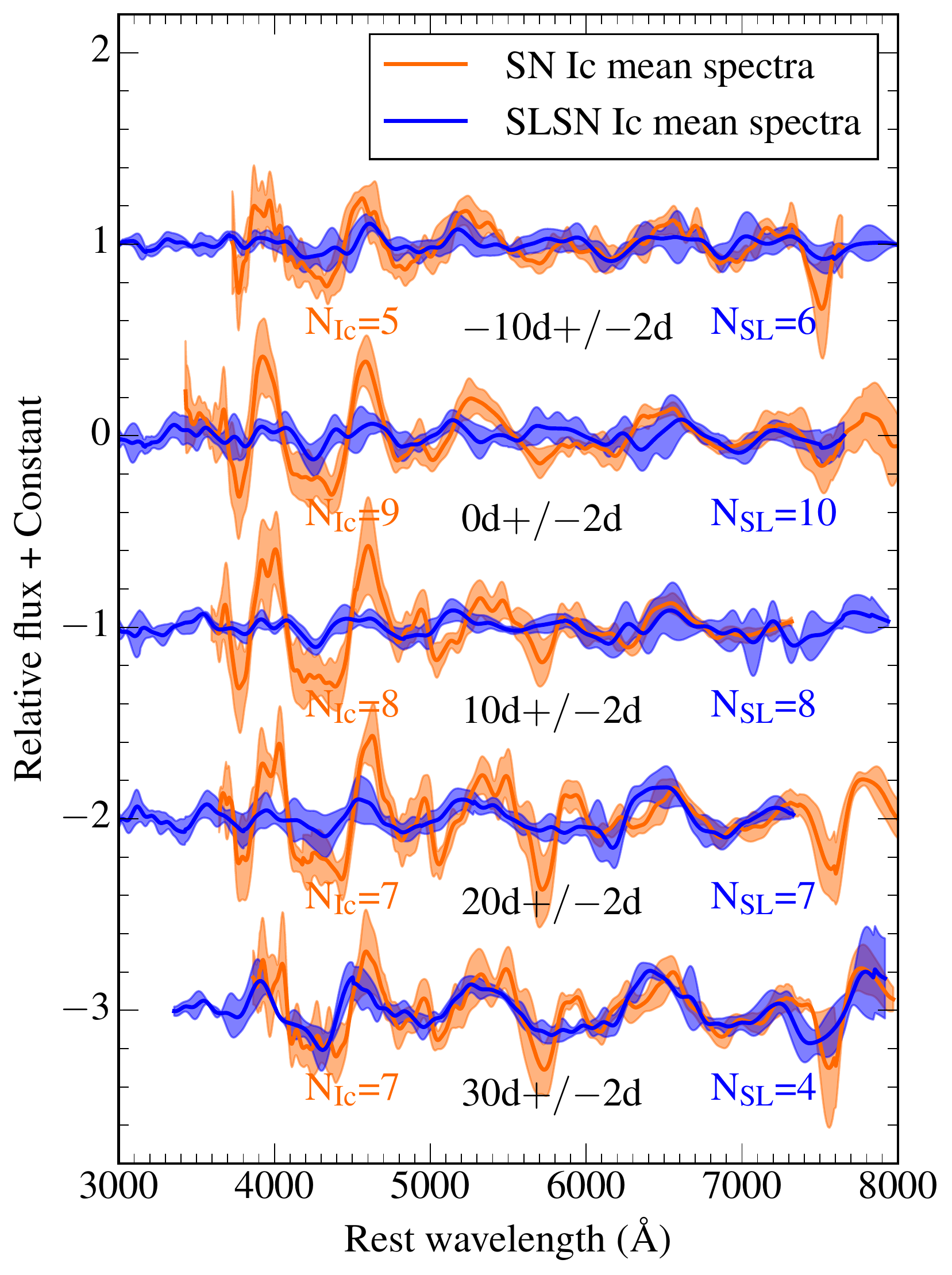}
}
\quad
\subfigure{%
\includegraphics[width=0.5\textwidth,scale=0.5,angle=0]{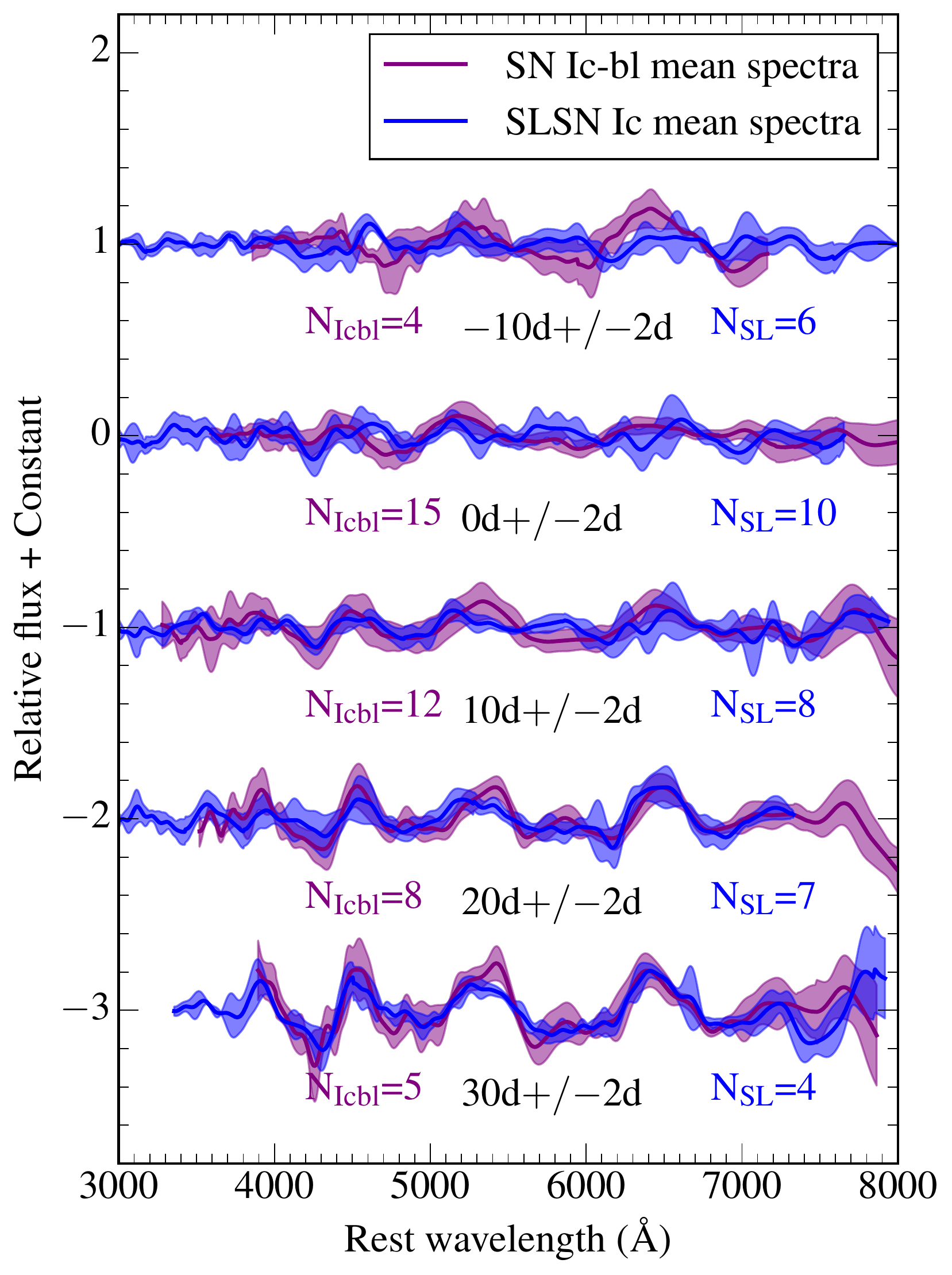}
}
\caption{Continuum-removed mean spectra and their corresponding standard deviations of SLSNe Ic (blue), SNe Ic-bl (green), and SNe Ic (red) at four different phase ranges: $t_{\mathrm{max}}=-12$ to $-8$, $-2$ to 2, 8 to 12, and 18 to 22 days. Each mean spectrum only includes one spectrum per SN even if multiple spectra have been taken within the phase range. N$_{\mathrm{SL}}$, N$_{\mathrm{Icbl}}$, and N$_{\mathrm{Ic}}$ represent the number of spectra (which is also the number of SNe) included in the mean spectrum of SLSNe Ic, SNe Ic-bl, and SNe Ic at each phase, respectively.}
\label{mean_spec}
\end{figure*}

\subsection{SLSNe Ic and SNe Ic-bl Have Comparable \FeII Velocities}
\label{velFeII}

\begin{deluxetable}{cccccc}
\tabletypesize{\scriptsize}
\tablecaption{Weighted mean absorption velocity of the \FeII and full width at half maximum (FWHM) of the convolved Gaussian kernel with the SN Ic template, at $t_{\mathrm{max}}\simeq10$ days\label{table_mean}}
\tablehead{
\colhead{SN type} &
\colhead{V$_{\mathrm{absorption}}$} &
\colhead{V$_{\mathrm{conv FWHM}}$ with respect to SNe Ic} \\
& (10$^3$ \kms) & (10$^3$ \kms)} \\
\startdata
SNe Ic &    $-8.0$ $\pm$ 1.4&       NA \\
SNe Ic-bl &   $-$18.5 $\pm$ 7.4 &       8.9 $\pm$ 2.1 \\
SLSNe Ic &   $-$15.0 $\pm$ 2.6 &       11.3 $\pm$ 3.6 
\enddata
\tablecomments{The errors are the weighted standard deviations of data that contribute to the weighted average value, which are indicated as the error bars in the figure that shows the weighted average values.}
\end{deluxetable}

\begin{figure*}[t]
\subfigure{%
\includegraphics[width=0.33\textwidth,scale=0.5,angle=0,trim = 0mm 0mm 0mm 0mm, clip]{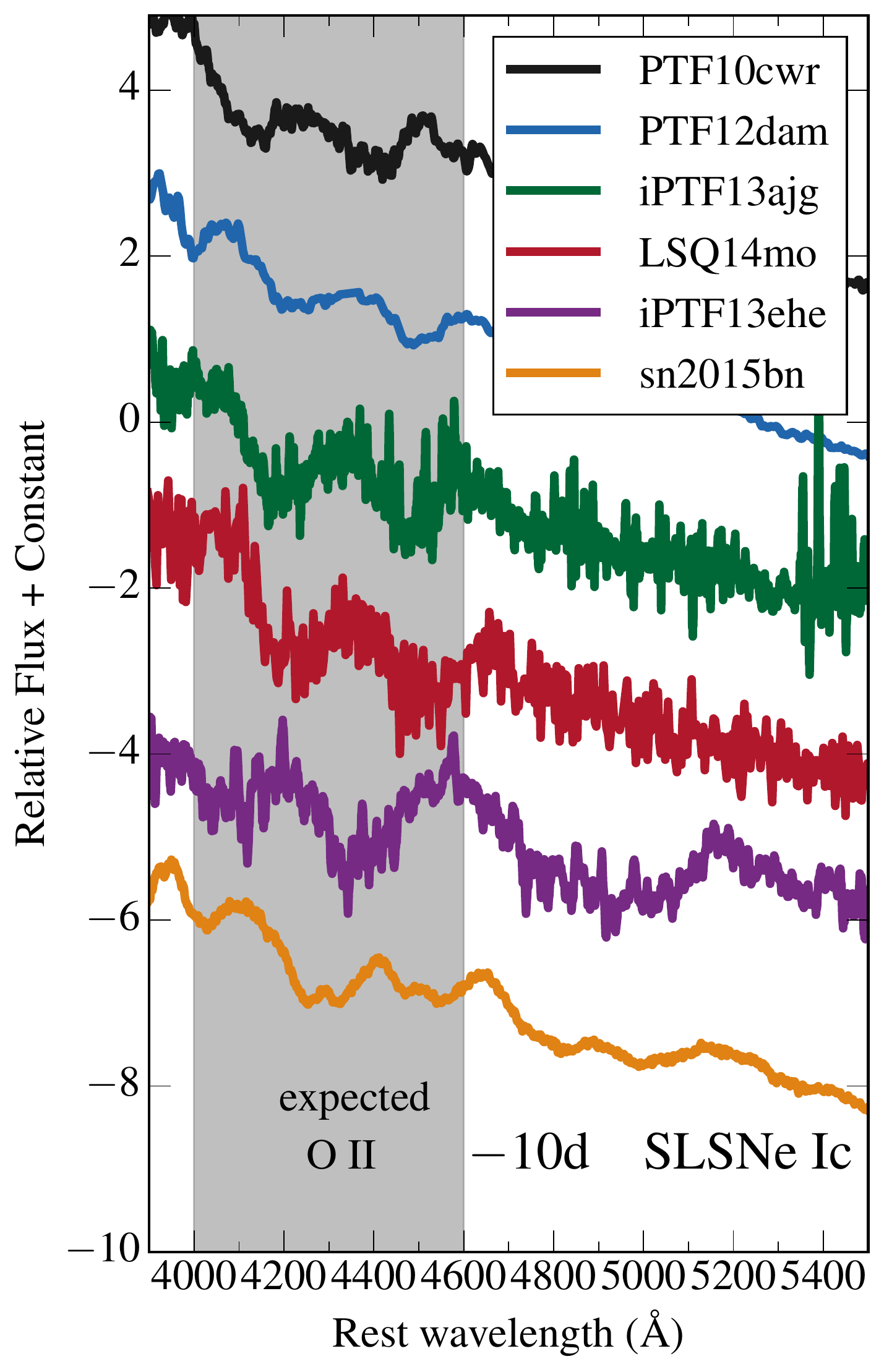}
}
\quad
\subfigure{%
\includegraphics[width=0.33\textwidth,scale=0.5,angle=0]{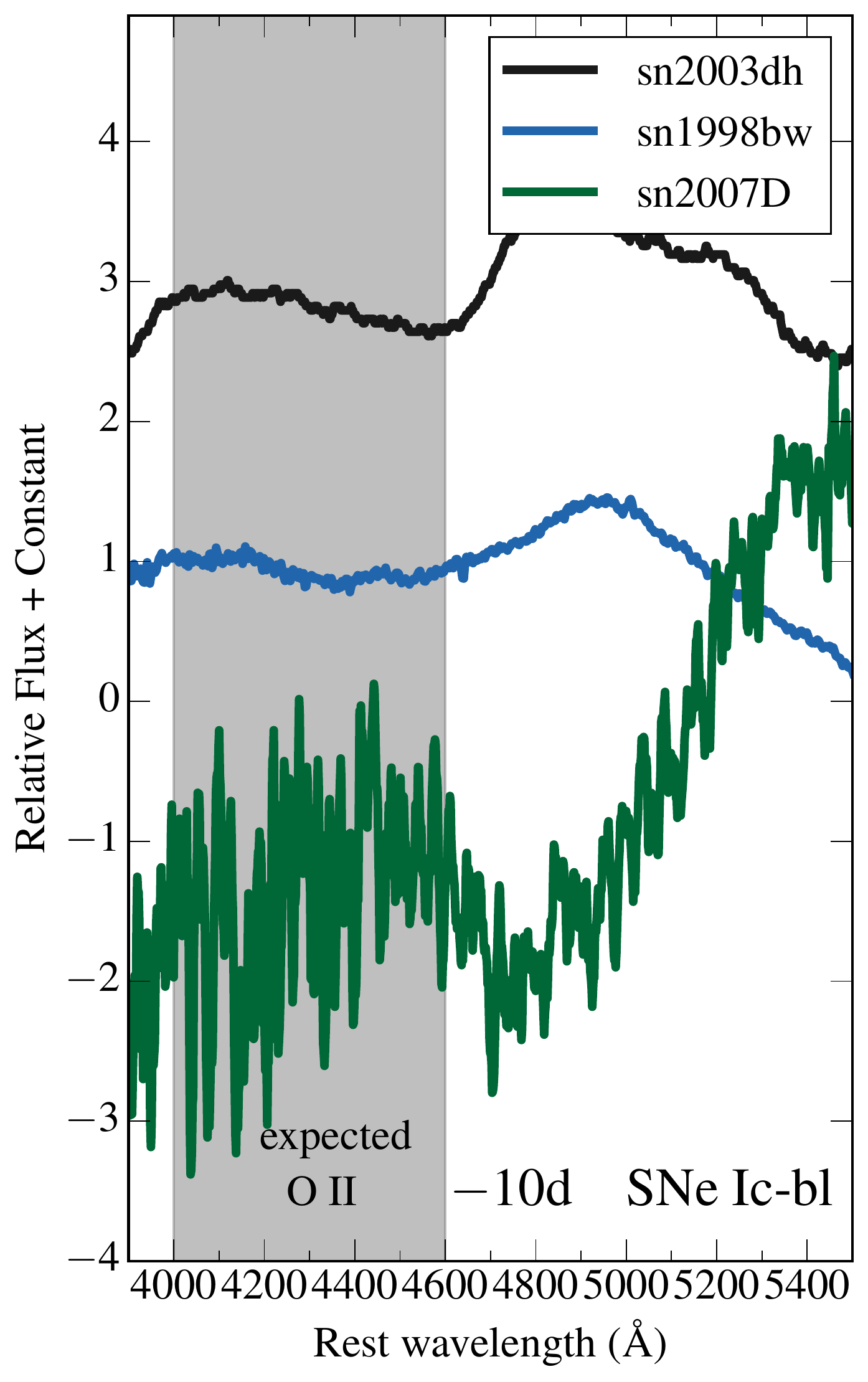}
}
\quad
\subfigure{%
\includegraphics[width=0.33\textwidth,scale=0.5,angle=0]{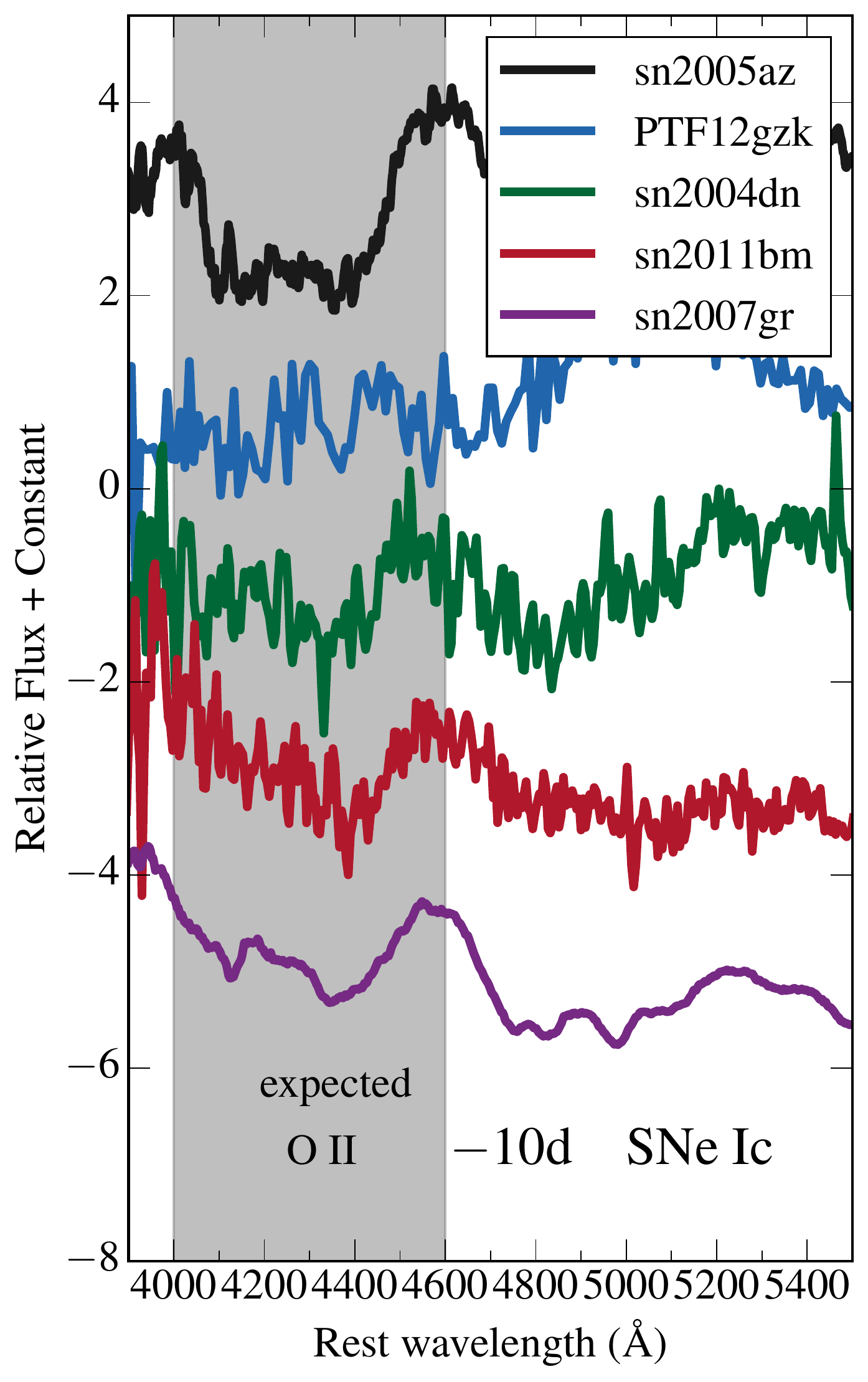}
}
\caption{Spectra of all SLSNe Ic (\textit{left}), SNe Ic-bl (\textit{middle}) and SNe Ic (\textit{right}) in our sample taken around $t_\mathrm{max}=-10$ days. The shaded area indicates the expected positions of the narrow `w' feature identified as O II lines, probably produced by the high temperatures at early times and non-thermal radiation \citep{mazzali16}. For display purpose, we smooth spectra of PTF10cwr, iPTF13ajg, iPTF13ehe and SN 2007D using a rolling average window of 10 data points. While there is no narrow `w' feature in SNe Ic-bl and SNe Ic, the narrow `w' feature in SLSNe Ic is likely due to O II lines, but for iPTF13ehe ($T_\mathrm{BB}\sim7$,000 K), the feature is identified as Mg II and Fe II lines in \citet{yan15}.}
\label{OII_mont}
\end{figure*}

We compared the \FeII velocities in SLSNe Ic, SNe Ic-bl and SNe Ic, since \FeII is a common line in these SN subtypes and has been suggested to trace the photospheric velocity (see footnote in Section \ref{vel_sum}). As discussed in Section \ref{vel},  we measured the \FeII velocities via template fitting as was done in \citet{modjaz15} to measure the velocities of \FeII in spectra of SNe Ic-bl. The absorption velocity evolution of the \FeII line for all individual SLSNe Ic in our sample is shown in the left panel of Figure \ref{plot_vel}, while the weighted average absorption velocities of \FeII for SLSNe Ic, SNe Ic-bl and SNe Ic are shown in the right panel. We calculated the average velocities as a Gaussian smoothed weighted average, where the weights correspond to the errors in the velocity measurements, and the Gaussian kernel has a standard deviation of 5 days. This provides a smoother velocity curve for each SN subtype than the method used in \citet{liu16} (their Appendix D) where the smoothing was achieved with a boxcar kernel (i.e., a rolling mean). In addition, in \citet{liu16} and \citet{modjaz15}, the measurements for each SN were first binned at 5-day granularity, while here each velocity measurement is used. Since Fe III lines may contaminate the \FeII line in SLSNe Ic at early phases, we disregarded measurements at $t_\mathrm{max}<10$ days when computing the weighted average velocity values for SLSNe Ic. 

There is a lot of diversity in \FeII absorption velocities for SLSNe Ic. For example, at $t_\mathrm{max}\simeq10$ days, PTF11rks has velocities around 20,000 \kms, while LSQ14mo has velocities around 10,000 \kms. However, we found no systematic difference in velocities for SLSNe Ic that had fast-declining light curves compared to those that had slow-declining light curves.\footnote{See definitions of fast-declining light curves and slow-declining light curves in \citet{inserra13}, \citet{nicholl16}, and \citet{Inserra17}.} The error bars on the weighted average velocities, which are calculated as the weighted standard deviations of the data and thus represent the diversity of the SNe velocities, are larger for SLSNe Ic and SNe Ic-bl than for SNe Ic, especially at early times. However, we note that error bars in all these SN subtypes are about 20\%-30\% of the absolute values. Moreover, the scatter in the weighted average measurement may be affected by the inaccurate measures of the date of maximum, since we have to use different filters, in which the date of maximum was measured, for different SLSNe Ic (see Table \ref{table} and Section \ref{diff_filter}). Thus, the intrinsic diversity in \FeII velocities of SLSNe Ic, SNe Ic-bl and SNe Ic could be smaller than that indicated by the error bars shown in the right panel of Figure \ref{plot_vel}. \citet{modjaz15} showed that although there is some overlap between \FeII absorption velocities in SNe Ic-bl and SNe Ic, the weighted average velocities in SNe Ic-bl are systematically higher than those in SNe Ic at almost all phases. Here, we have found similar patterns in \FeII absorption velocities between SLSNe Ic and SNe Ic. Although the \FeII velocities of individual SLSNe Ic are not clearly separated from velocities of individual SNe Ic, as a population, SLSNe Ic have \FeII velocities that are higher on average than those in SNe Ic at almost all phases and are similar to those in SNe Ic-bl. In particular, as shown in Table \ref{table_mean}, the weighted average \FeII absorption velocities of SLSNe Ic and SNe Ic-bl are about two times that of SNe Ic at $t_\mathrm{max}=10$ days. While we can measure the absorption velocity of this line in SLSNe Ic only starting at $t_\mathrm{max}=10$ days, which at that phase has an average value of 15,000 ($+/-$ 2,600) \kms, we note that this implies that the absorption velocity of SLSNe Ic at the date of maximum light, which is usually the one used for ejecta mass estimates, will have to be even higher. Moreover, \citet{jerkstrand16} found that SLSNe Ic and SNe Ic-bl have similar nebular-phase spectra in terms of velocities. These similarities in observations indicate that SLSNe Ic and SNe Ic-bl may have similar explosion engines, which is consistent with a multi-dimensional magnetar model in \citet{Suzuki16}, where they claimed that SLSNe Ic and SNe Ic-bl could be produced by a similar engine with different energy injection rates. If SLSNe Ic and SNe Ic-bl also have similar ejecta masses (which may not be the case, e.g, \citealt{nicholl15}), then they may have similar kinetic energies, given they have broadly similar photospheric velocities as traced by \FeII line.  

We note that weighted average velocities of \FeII in SLSNe Ic, SNe Ic-bl and SNe Ic converge to $\sim7,000$ \kms~starting at $t_\mathrm{max}=35$ days. Thus, when we compare absorption velocities of different subtypes, we focus on measurements taken at $t_\mathrm{max}<35$ days.

\subsection{SLSNe Ic and SNe Ic-bl Have Comparable Spectral Features}
\label{AveSpecSL}

We compared spectral features in SLSNe Ic, SNe Ic-bl and SNe Ic by comparing continuum-removed average spectra at different phases. We divided the continuum out from raw spectra in order to remove the impact of any potential reddening caused by dust and to focus on spectral features. The left panel of Figure \ref{mean_spec} compares continuum-removed average spectra of SLSNe Ic with those of SNe Ic, while the right panel compares SLSNe Ic with SNe Ic-bl. Obviously, SNe Ic have stronger and narrower features than SLSNe Ic at many wavelengths. In particular, from 4000 \AA~to 6000 \AA~at $t_\mathrm{max}=0$, 10, and 20 days, SLSNe Ic mean spectra are up to 2 standard deviations away from the SNe Ic mean spectra. As listed in Table \ref{table_mean}, at $t_\mathrm{max}=10$ days, the average full width at half maximum (FWHM) of the Gaussian kernel convolved with the SN Ic template is around 11,000 \kms. In contrast, SLSNe Ic and SNe Ic-bl have similar features in terms of width and strength at most wavelengths, although there are mismatches around 4600 \AA~and 6000 \AA~at $t_\mathrm{max}=-10$ and 0 days, as well as around 4300 \AA~at $t_\mathrm{max}=-10$ as we discuss in detail in Section \ref{OII}. We checked that the broad lines are present in individual SLSN Ic spectra, and thus the broad lines in average spectra are not an artifact of the averaging process. We note that although there is no clear separation between SLSNe Ic (or SNe Ic-bl) and SNe Ic in terms of width of \FeIIn, SLSNe Ic and SNe Ic-bl have on average systematically broader \FeII lines than SNe Ic (Table 3). The fact that the mean (and standard deviation) spectra of SLSNe Ic and SNe Ic-bl overlap at many wavelengths in spectra at photospheric phases again indicates that SLSNe Ic and SNe Ic-bl may have similar dynamics. This is important because SLSNe Ic are usually compared to SNe Ic \citep[e.g.,][]{Pastorello10, inserra13}. Actually, the average spectrum of SLSNe Ic at $t_\mathrm{max}=-10$ days is similar to the average spectrum of SNe Ic at $t_\mathrm{max}=30$ days, thus, we have partly validated the claim in \citet{Pastorello10} and \citet{inserra13}, namely that SLSN Ic spectra look similar to SNe Ic with a phase lag of about one month. However, more importantly, at the same phase, we have found that certain spectral characteristics (such as line widths and \FeII velocities) of SLSNe Ic are more similar to those of SNe Ic-bl than those of SNe Ic. We note that a large difference between SLSNe Ic and SNe Ic-bl spectra is that the continuum in SLSNe Ic is much bluer than in both SNe Ic-bl and SNe Ic (not shown in Figure \ref{mean_spec}).

\subsection{O II Feature Around 4300 \AA~Can be Used to Identify SLSNe Ic at Early Time}
\label{OII}

The narrow `w' feature around 4300 \AA~in SLSN Ic spectra before $t_\mathrm{max}=0$ day is observed and identified as O II lines \citep[even sometimes seen as a number of distinct features;][]{quimby11, gal-yam16} in the literature \citep{quimby07_05ap, quimby11, gal-yam12, nicholl15, mazzali16, gal-yam16}. We have checked the spectra of all 32 SLSNe Ic in our sample, and found that all SLSNe Ic have a narrow `w' feature around 4300 \AA, except for 12 SLSNe Ic that do not have spectra taken before the date of maximum light, 4 SLSNe Ic that do not have spectra covering the 4300 \AA~region, and 1 SLSN Ic (SN 2011kl) with spectrum that are too noisy. We have also checked the spectra of all 23 SNe Ic-bl and 17 SNe Ic in our sample. There are 18 SNe Ic-bl and 13 SNe Ic that have spectra before $t_\mathrm{max}=0$ day and cover the 4300 \AA~region. However, none of them show a similar narrow `w' feature as seen in SLSN Ic spectra.\footnote{The very early-time spectra of SN Ib 2008D did show the same `w' feature in one spectrum at 1.8 days after the shock-breakout or 16.6 days before the V-band maximum, which may have been produced at high temperatures \citep{modjaz09}.}

We show spectra of all SLSNe Ic, SNe Ic-bl and SNe Ic in our sample that were taken around $t_\mathrm{max}=-10$ days in Figure \ref{OII_mont}. The shaded area indicates the expected positions of the narrow `w' feature due to O II lines produced at high temperatures and via non-thermal excitation \citep{mazzali16}. Spectra of SLSNe Ic show the narrow `w' feature between $\sim4000$ \AA~and $\sim4600$ \AA, while spectra of SNe Ic-bl and SNe Ic do not show such narrow `w' feature within a similar wavelength range. Thus, we suggest to use the narrow `w' feature around 4300 \AA, along with other spectral and photometric properties, to identify SLSNe Ic at early time, i.e., before maximum light, as initially proposed by \citet{quimby11} based on a smaller sample. We note that not all narrow `w' features in SLSNe Ic before the peak epoch are identified as O II lines. For example, a low temperature SLSN Ic iPTF13ehe ($T_\mathrm{BB}\sim7$,000 K) shows a narrow `w' feature at $t_\mathrm{max}=-10$ days, but is identified as due to Mg II and Fe II lines \citep{yan15}. Except for iPTF13ehe, the narrow `w' feature in other SLSNe Ic is likely due to blended O II lines \citep[see table 2 in][]{mazzali16}, which may make it hard to measure the absorption velocity of the feature. 
 
%
%
%
\section{Comparisons of Measured Velocities with Model Predictions}
\label{comp_model}

\subsection{Tension between Our Measured High Velocities and the Interaction Model}
\label{highVel}

In the interaction model for SLSNe, the SN ejecta interacts with circumstellar material that was ejected before the explosion, thus converting the large kinetic energy of the ejecta into radiation (e.g., \citealt{ofek14} and references therein). While this model can reproduce SLSNe IIn well (i.e., those with hydrogen emission lines; \citealt{smith10}), this model has a number of challenges for SLSNe Ic: the circumstellar material would have to be completely free of both hydrogen and helium (if helium were present, one would expect to observe helium emission lines,\footnote{We note that SN 2012il appears to show He I $\lambda$10830 in emission in a spectrum at $t_{\mathrm{max}}=53$ days, but none of the optical He I lines in a spectrum taken at the same epoch \citep{inserra13}, neither in absorption, nor in emission, which is odd and has never been observed before in any SN.} as seen in SNe Ib-n, e.g., \citealt{foley07,pastorello08,pastorello16,shivvers16}). However, even for hydrogen-free and helium-free material one would still expect to see narrow lines of oxygen and other elements (e.g., as observed by \citealt{benami14} in SN 2010mb), which are not seen in any SLSNe Ic, though another explanation may be that the abundances or ionization/excitation conditions are not sufficient to give rise to them. In addition, the observed large absorption velocities have not been reproduced by the interaction model which at the same time has to reproduce the observed broad light curves \citep{sorokina16}. \citet{sorokina16} speculate that the observed spectra of SLSNe Ic could be explained by interaction with a separate shell that is expanding rapidly with a peculiar velocity structure. That velocity structure would be such that the inner radius of the shell is almost at zero velocity, while the outer radius is at the velocity observed in absorption in the spectra. However, since we find that the absorption velocities of SLSNe Ic (see Section \ref{velFeII}) are even higher than the previously adopted velocity of 10,000 \kms~for constraining models \citep{sorokina16}, there is even more tension - more importantly, we note the SLSN Ic spectra have also broad spectral features (see Section \ref{AveSpecSL}), indicating a relatively large mass at a high velocity, not just a thin layer, as would be the case in the shell model. Alternatively, as was suggested for the SN Ia 2009dc by \citet{hachinger12_Ia}, the interaction layer may not be responsible for the absorption lines, but only lead to their dilution, and only contribute to the continuum in the blue. In any case, we suggest that future works, which attempt to explore interaction models for SLSNe Ic, employ sophisticated hydrodynamical models and synthesize spectra at higher resolution than done in \citet{sorokina16} in order to try to reproduce the average spectra of the full SLSNe Ic population.

\subsection{Testing the One-dimensional Magnetar Model using Flat Velocity-evolution of \FeII}
\label{FlatVel}

\begin{figure}[t]
\includegraphics[scale=0.5,angle=0]{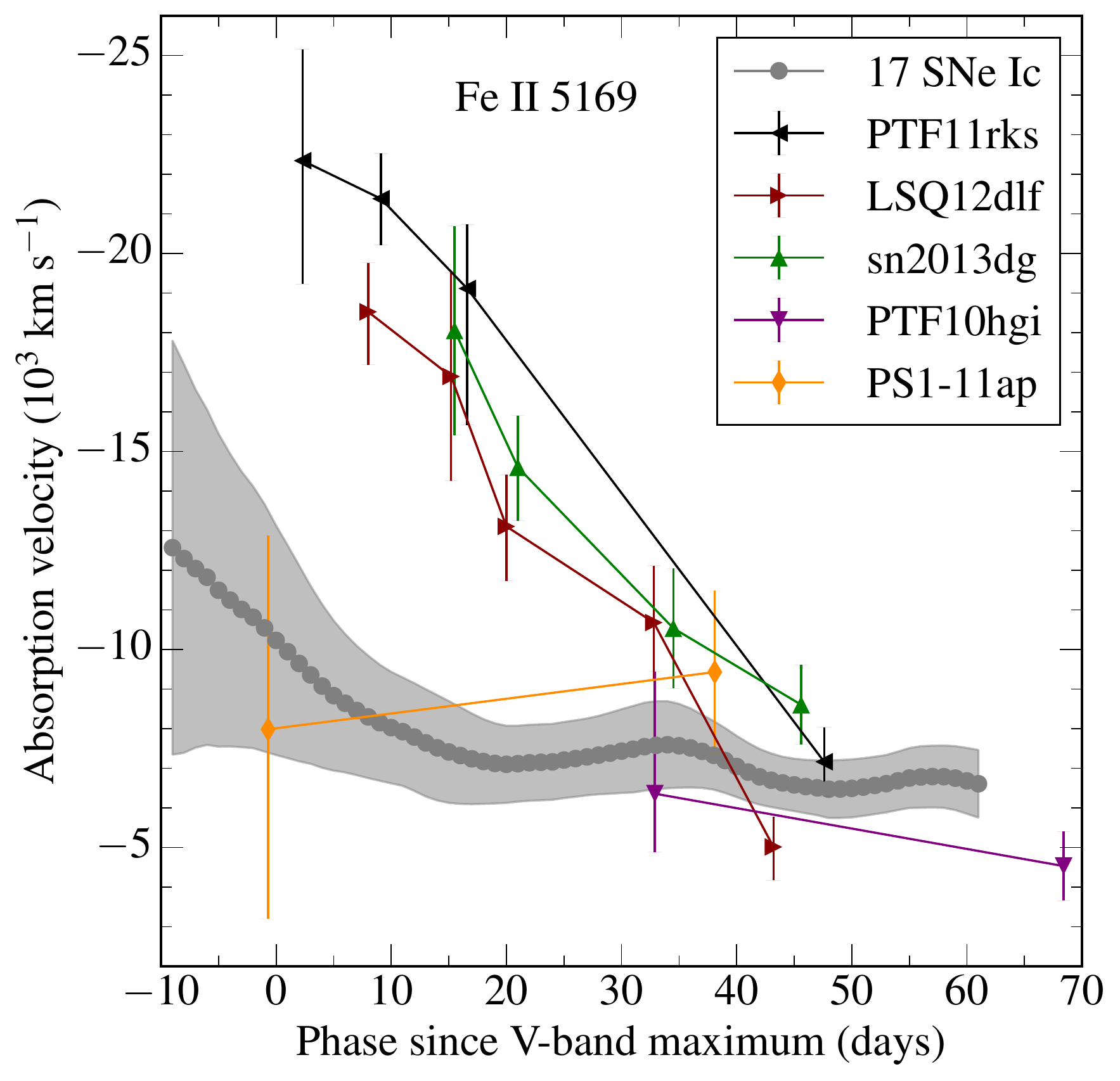}
\caption{Measured \FeII velocities for all slow-evolution and fast-evolution velocity SLSNe Ic, which are used to check the predictions of the one-dimensional magnetar model and satisfy the following: spectrum phases cover more than 10 days; the velocity change with time is either less than 2,000 \kms~or larger than 8,000 \kms. In contrast, weighted average \FeII velocities of SNe Ic in Figure \ref{plot_vel} are shown in gray.}
\label{flat_vel}
\end{figure}

One promising power source model for SLSNe Ic is the magnetar model \citep[e.g.,][]{kasen10, woosley10, inserra13, nicholl13, Metzger15, chen16, Suzuki16}. In the one-dimensional magnetar model, there will be a mass shell formed due to the magnetar bubble. Based on radiation hydrodynamic calculations, the mass shell is supposed to cause an observational imprint via a prolonged period of flat velocity-evolution, i.e., a velocity plateau, in the observed spectra \citep{kasen10}. Some observational papers claim that they have detected such a velocity plateau in their analyses of SLSN Ic spectra. For example, the velocity plateau is claimed in PS1-10ky and PS1-10awh using lines of C II $\lambda$2330, Si III $\lambda$2540, and Mg II $\lambda$2800 \citep{chomiuk11}. \citet{nicholl14} reported velocity plateaus in SN 2013dg using Mg II $\lambda$4481 and Mg I] $\lambda$4571. \citet{nicholl15} have suggested that the \FeII velocity of the following SLSNe Ic has a flat evolution as a function of time: PTF11rks, PTF12dam, SN 2013dg, and LSQ14mo.

\begin{figure*}[t]
\subfigure{%
\includegraphics[scale=0.4,angle=0]{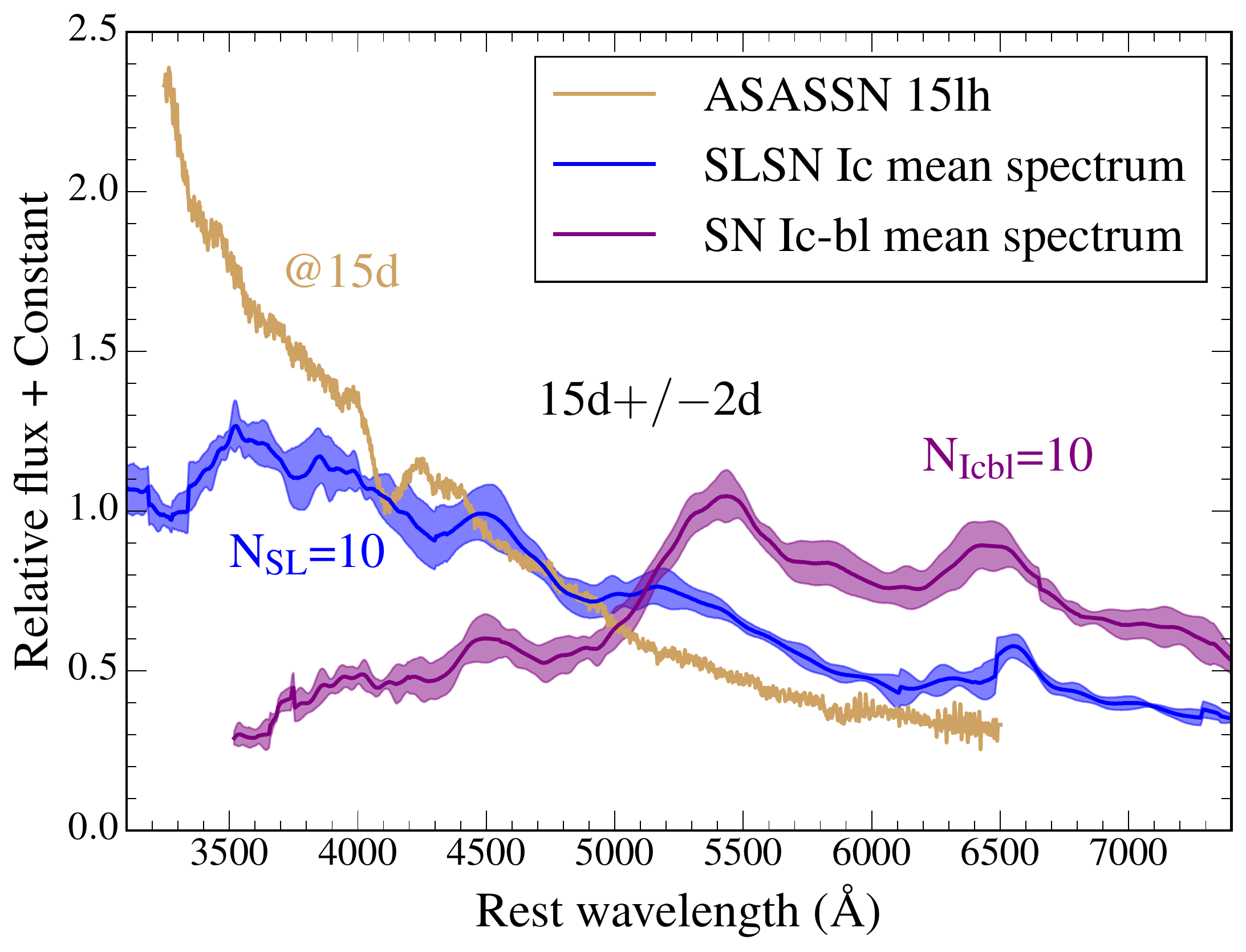}
}
\quad
\subfigure{%
\includegraphics[scale=0.4,angle=0]{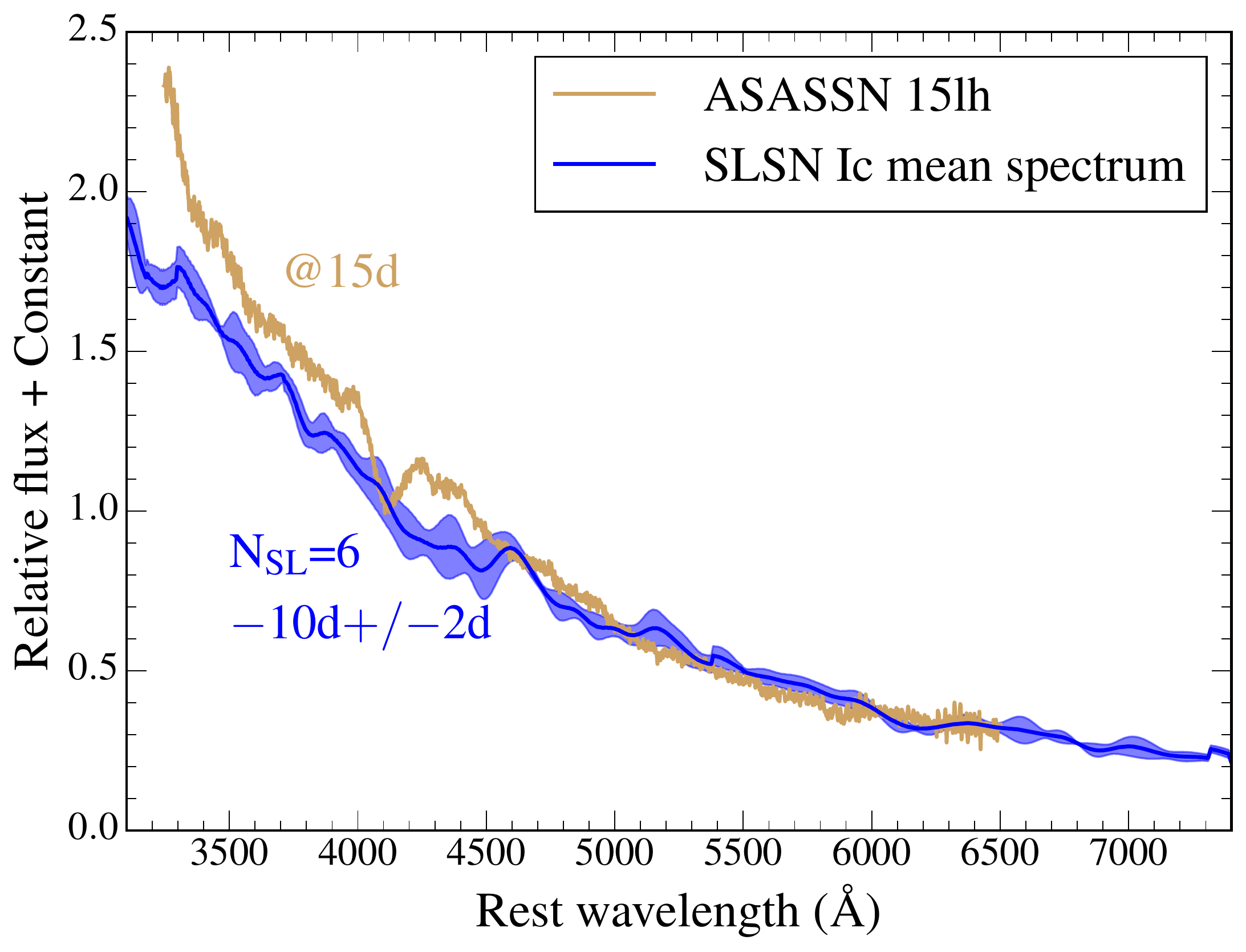}
}
\caption{Spectrum of ASASSN-15lh (yellow) at $t_{\mathrm{max}}=15$ days in comparison with mean spectra of different SN types in order to probe whether this transient was a SN. Mean spectra and their corresponding standard deviations of SLSNe Ic (blue) and SNe Ic-bl (green) at $t_{\mathrm{max}}=15\pm2$ days (\textit{left}) and $t_{\mathrm{max}}=-10\pm2$ days (\textit{right}). Each mean spectrum only includes one spectrum per SN even if multiple spectra have been taken within the phase range. N$_{\mathrm{SL}}$ and N$_{\mathrm{Icbl}}$ represent the number of spectra (which is also the number of SNe) included in the mean spectrum of SLSNe Ic and SNe Ic-bl at each phase, respectively.}
\label{mean_15lh}
\end{figure*}

In our work, we have explored the velocity evolution of \FeII using our own measurements that utilizes our novel technique (see Section \ref{method}). First, we define flat (or slow) velocity-evolution SNe as SNe whose spectra span more than 10 days and for which the \FeII velocities decrease by less than 2,000 \kms. With this definition, we only identify two slow velocity-evolution SLSNe Ic (PTF10hgi and PS1-11ap) out of 13 SLSNe Ic that satisfy the phase requirement and plot them in Figure \ref{flat_vel}. In contrast, we also show fast velocity-evolution SLSNe Ic (PTF11rks, LSQ12dlf, and SN 2013dg) whose spectra cover more than 10 days, but whose velocities decrease by more than 8,000 \kms. Changing the requirement for the time range over which the SLSNe Ic should have spectra from 10 days to a more lenient 5 days does not change our conclusions, since the only SLSNe Ic fulfilling the new requirement are the same objects as before. The slow-evolution SLSNe Ic are consistent with a dense shell model, while the fast-evolution SLSNe Ic are consistent with a simple spherical model. A dense shell can be formed in a magnetar model but there are other channels, e.g., via eruptions before SNe explode. We note that even for PTF10hgi and PS1-11ap, it is not clear that their velocity evolution should be used to argue for the one-dimensional magnetar model: PTF10hgi has a flat evolution in velocity, but like SNe Ic that show velocity plateaus at similar phases, this could be just indicating that the flat evolution is due to radiative transfer effects and not reflect a physical ejecta structure; PS1-11ap has a large error bar for the first data point, making it consistent with a high velocity value, and thus no flat evolution.

In the following, we compare \FeII velocities measured in this work with those in \citet{nicholl15} for 7 SLSNe Ic we both have in common, and most of our measurements are consistent within the error bars. However, there are different systematic trends that lead to qualitatively different conclusions for both works. For 4 SLSNe Ic (PTF11rks, PTF12dam, SN 2013dg, and LSQ14mo), we do not find a flat velocity evolution for the \FeII line as claimed by \citet{nicholl15}. One reason is that we have measurements for the earliest and latest spectra that are not presented by \citet{nicholl15}. For PTF11rks, we have a \FeII velocity at $t_\mathrm{max}\simeq50$ days, which is much lower than velocities at $t_\mathrm{max}<20$ days, while \citet{nicholl15} only presented \FeII velocities before $t_\mathrm{max}=20$ days. A similar reason applies to PTF12dam as well. Another reason is that the way we measure the \FeII velocity is different from that in \citet{nicholl15}. We fit the \FeII region using the similar region in broadened and blue-shifted SN Ic spectra in a MCMC framework, while \citet{nicholl15} fit the whole \FeII region (which for SLSNe Ic is a blend and includes 2 additional Fe II lines, namely Fe II $\lambda\lambda$4924, 5018) with only one Gaussian profile. As discussed in Section \ref{velComp}, although both methods have advantages and disadvantages, the template fitting method in our work can better capture the \FeII line when it is blended with the other two Fe II lines as shown in \citet{modjaz15} for SNe Ic-bl.  

We note that a flat evolution in velocities is observed in SNe other than SLSNe Ic. In particular, the right panel of Figure \ref{plot_vel} shows that the average velocity evolution of SNe Ic becomes flat after $t_\mathrm{max}=20$ days. The velocity plateau of He I lines in SNe Ib (SN 2006el, SN 2009mg, and SN 2011dh) and SNe IIb (SN 1998dt, SN 1999ex, SN 2005bf, and SN 2007Y) is also observed by \citet{folatelli14} and \citet{liu16}. However, the magnetar model is unlikely to be the reason for the velocity plateaus in these stripped SNe. Thus, we suggest that any claim of the magnetar model based on the flat velocity-evolution of a line observed in spectra of SLSNe Ic has to be checked by comparing to the velocity-evolution of the same line in normal SNe Ic over the same time period. If a line exhibits a velocity plateau in both normal SNe and in SLSNe Ic over the same time period, then it may be just due to radiative transfer effects, and thus should not be used as a diagnostic for the central power source.

We conclude that as a whole population, the velocity evolution as traced by \FeII of SLSNe Ic is not consistent with that predicted by the one-dimensional magnetar model \citep{kasen10}, while individual SLSNe Ic might be. The obvious next step of performing magnetar calculations in two dimensions has been recently tackled. \citet{chen16} and \citep{Suzuki16} found that the dense shell which is commonly seen in one-dimensional simulations and which accounts for the predicted velocity plateau, is destroyed by multi-dimensional effects, including turbulence. Thus, while we await spectral synthesis calculations based on two-dimensional models for their predicted velocity evolution, it is reasonable to expect that two-dimensional models will not show such a velocity plateau and thus, will be more in line with our observational results. Moreover, \citet{inserra13} showed that a semi-analytical diffusion model in the magnetar scenario can reproduce light curves of SLSNe Ic well, but the predicted photospheric velocity evolution is not flat.

\subsection{Caveats when Comparing Observations to One-dimensional Magnetar Model}
\label{caveat}

In our analysis we have used \FeII to trace the velocity structure of the SN Ic family. While using this line to perform inter-comparisons between different SN subtypes (Section \ref{velFeII}) is valid and indeed required, since different lines have systematically different velocity structures \citep{modjaz15}, we now note some of the caveats when using that line to constrain theoretical models. Theoretical models assume a `photospheric velocity' and the question is which observed line traces it best. On the one hand, \citet{branch02} suggest that \FeII traces well the photospheric velocity based on their SYNOW models for stripped-envelope core-collapse SNe (stripped SNe). On the other hand, \citet{dessart15} conduct a careful exploration using progenitor, explosion and non-LTE radiative transfer models for stripped SNe and find that while the photosphere, and therefore its associated velocity, is poorly defined, the O I $\lambda$7774 velocity is a good indicator of the representative ejecta velocity for SNe Ib and Ic at the date of maximum light. We showed in \citet{liu16} that while the absolute values of the \FeII and O I $\lambda$7774 velocities are somewhat different for SNe Ib and Ic, the systematic offset between the 2 subtype velocities was shown by both lines, i.e., the relative velocity values are correct. In this work, we did not use O I $\lambda$7774 as an indicator for photospheric velocity since most of our SLSN Ic spectra do not cover its wavelength range.

Thus, our conclusions about the one-dimensional magnetar model in the preceding section may not stand if another line were to be used, since different lines have different velocity behavior (e.g, Si II $\lambda$6355 vs. \FeIIn, see Figure 4 in \citealt{modjaz15}). However, as we note in Section \ref{FlatVel}, the same line needs to be used for both SNe Ic and SLSNe Ic to ensure that any plateau behavior seen in the velocities of SLSNe Ic is not due to radiative transfer effects, which would be reflected in the same plateau behavior for ordinary SNe Ic.

From the theoretical point of view, a few assumptions have to be adopted for the comparison between theory and observations. For specifically the one-dimensional magnetar model, we used the following prediction based on Equation 7 and Figure 2 in \citet{kasen10}: if the equation predicts a shell velocity of 10,000 \kms, then based on the figure, the velocity plateau will start 10 days after the explosion. Of course if the magnetar energy were to be lower by a factor of 12--13, the phase range of the velocity plateaus would be at 100 days after the explosion, a regime not probed by the optical spectra in our sample. 

\subsection{Caveats if SLSNe Ic are explosions with large-scale asphericities that affect their luminosity}
\label{asymmetry}
Determining the geometry of SLSNe Ic explosions is crucial for differentiating between the different powering mechanisms. For example, an explosion with a central engine may be an intrinsically asymmetric explosion and would leave imprints on polarimetric observations. As discussed in Section \ref{data}, due to the lack of observations, it is hard to conclude whether all SLSNe Ic are non-spherical explosions. Indeed, our observations disfavor the spherically symmetric magnetar model in one-dimension (see Section \ref{FlatVel}), though this does not necessarily mean that SLSNe Ic have to harbor large-scale asphericities. If SLSNe Ic are powered by highly aspherical engines, then one may speculate that the discovered SLSNe Ic appear luminous because they are seen on-axis, and thus the SLSNe Ic in our sample would be biased towards high luminosity explosions. Even if this conjecture entails that the population properties of SLSNe Ic may be biased, the velocity measurements of the individual SNe would still be valid, as well as our discussions in Sections \ref{highVel} and \ref{FlatVel}, except that we would have to compare the velocities of SLSNe Ic to those of SNe Ic-bl with observed GRBs only, not the whole population of SNe Ic-bl, a change that would not change our conclusions.

\section{Special SLSNe Ic}
\label{weird_spec}

\subsection{Spectrum of the most luminous `SN' ASASSN-15lh does not resemble the average SLSNe Ic spectra}

\citet{dong16} claimed ASASSN-15lh (AKA SN 2015L) as the most luminous SN ever discovered. They found that the spectra of ASASSN-15lh are blue and featureless except for a deep and broad trough near 4100 \AA. They also found that the trough in ASASSN-15lh resembles O II feature in PTF10cwr (AKA SN 2010gx). However, whether ASASSN-15lh is a SN or not is still debated. On the one hand, \citet{leloudas16} found that their observations are more consistent with a tidal disruption event (TDE), based on temperature evolution, transient location in the host galaxy, and other evidence. They also found that if the trough in ASASSN-15lh were due to O II lines, there would need to be an additional strong feature around 4400 \AA, which is not observed in the spectra of ASASSN-15lh. Moreover, \citet{Margutti16} favored a TDE origin for ASASSN-15lh based on analysis of X-rays from the host galaxy, although more monitoring of the X-rays is needed to truly distinguish between the two origins. On the other hand, spectra of ASASSN-15lh lack both hydrogen and helium emission lines, something that has not been observed in TDE spectra so far \citep[e.g.,][]{arcavi14}. The host galaxy does not resemble the typical hosts of SLSNe \citep{Godoy-Rivera16} and is too massive for a TDE \citep{dong16}. Moreover, the rebrightening in UV is also unexpected for both SLSNe and TDEs \citep{Godoy-Rivera16, brown16}.

In our work, we partly address the question of whether ASASSN-15lh is a SN or not by comparing its spectra to average spectra (with continua) of SLSNe Ic and SNe Ic-bl. In Figure \ref{mean_15lh}, we compare the spectrum of ASASSN-15lh at $t_\mathrm{max}=15$ days to average spectra and their corresponding standard deviations of SLSNe Ic at $t_\mathrm{max}=15$, and $-10$ days, as well as to the average spectrum of SNe Ic-bl at $t_\mathrm{max}=15$ days. The left panel shows that there is no similar feature in average spectra of SLSNe Ic and SNe Ic-bl as the trough in ASASSN-15lh at $t_\mathrm{max}=15$. Moreover, the continuum of ASASSN-15lh does not resemble those in average spectra of SLSNe Ic at short wavelengths and SNe Ic-bl at optical wavelengths. The right panel shows the same ASASSN-15lh spectrum at $t_\mathrm{max}=15$ days and the SLSN Ic average spectrum and its corresponding standard deviations at $t_\mathrm{max}=-10$ days. While the trough in ASASSN-15lh is still not consistent with features in the average spectrum within 1 standard deviation, the continua in both spectra are similar. We have several more spectra of ASASSN-15lh at later phases until $t_\mathrm{max}=39$ days. However, they all resemble the spectrum at $t_\mathrm{max}=15$ days in continuum and spectral features. Thus, even if ASASSN-15lh were a SLSN I, it would be a very slowly evolving SLSN Ic with peculiar features. 

\subsection{The first SN-ULGRB connection SN 2011kl / GRB 111209A}

\begin{figure}[t]
\includegraphics[scale=0.4,angle=0]{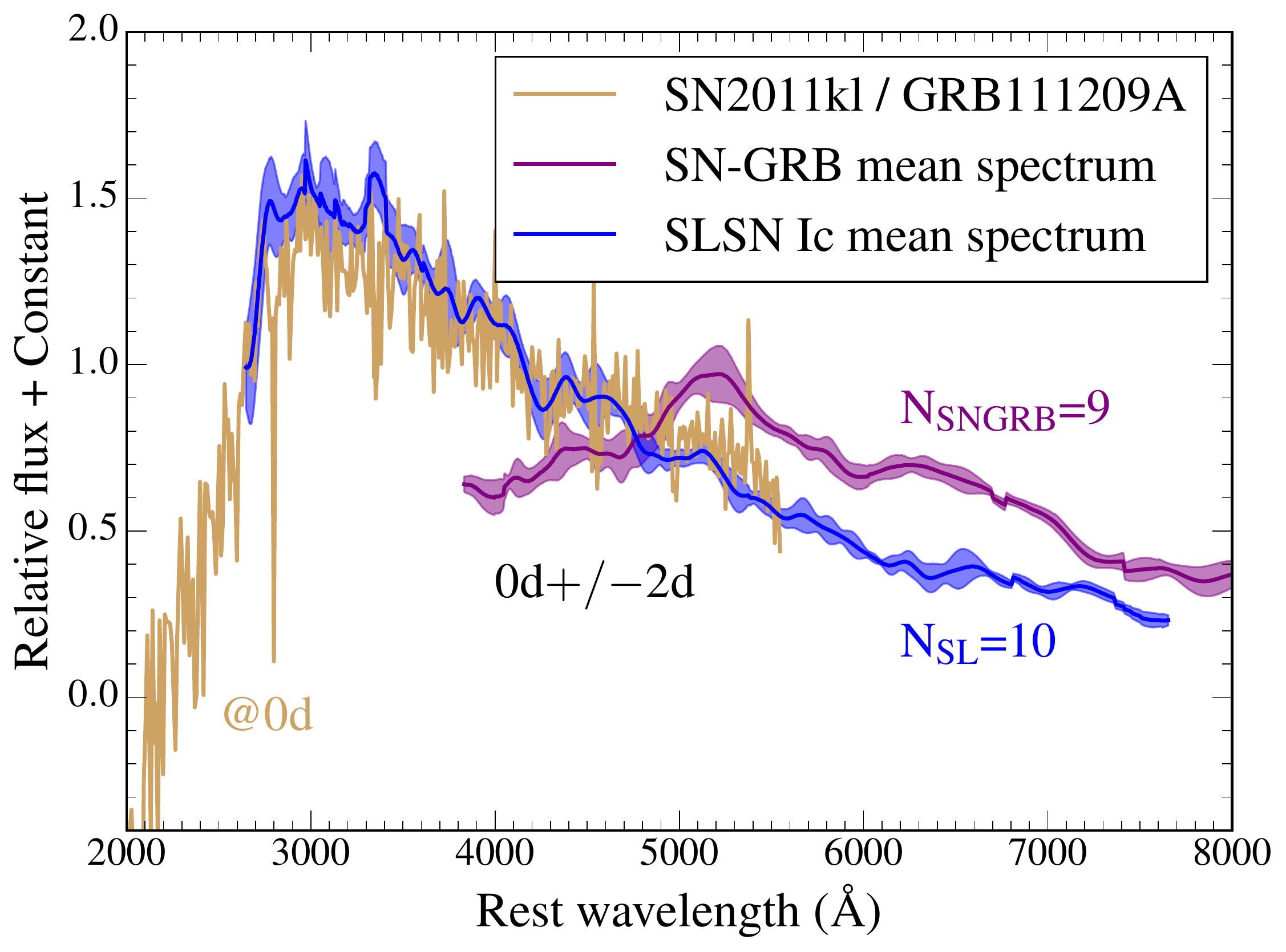}
\caption{Same as Figure \ref{mean_15lh}, but for spectrum of SN 2011kl at $t_{\mathrm{max}}=0$ day, as well as mean SN spectra of SN-GRBs and SLSNe Ic at $t_{\mathrm{max}}=0\pm2$ day.}
\label{mean_11kl}
\end{figure}

SN 2011kl is another unusual SLSN Ic since it constitutes the first detection of a SN explosion associated with an ultra-long duration gamma ray burst (GRB), namely ULGRB 111209A \citep{Greiner15, Kann16}. SN 2011kl is more luminous than SNe Ic-bl associated with long-duration GRBs (SN-GRBs; for recent review, see e.g., \citealt{modjaz11, Hjorth12, cano16}), but not as luminous as most, if not all, SLSNe Ic \citep{Greiner15, Kann16}. In Figure \ref{mean_11kl}, we compare the spectrum of SN 2011kl to the average and standard deviation spectra of SN-GRBs and SLSNe Ic at similar phases. Note that we could not measure the \FeII velocity in SN 2011kl due to the noisy spectra. Thus, we did not compare the \FeII velocities. We find that the continuum of SN 2011kl's spectrum does not resemble those of the population of SN-GRBs, but is fully consistent with those in the population of SLNe I; however, we find the spectrum of SN 2011kl to be too noisy to robustly evaluate whether the widths of the line features are as broad as in other SLSNe Ic and SN-GRBs. While \citet{Greiner15} came to a similar conclusion based on fewer objects in their comparison set (only 3 SLSNe Ic and 1 SN-GRB), we have conducted this spectroscopic comparison with the full populations of SLSNe Ic and SN-GRBs. Thus, SN 2011kl could be a true SLSN I. 

\subsection{SN 2007bi}

\begin{figure}[t]
\includegraphics[scale=0.4,angle=0]{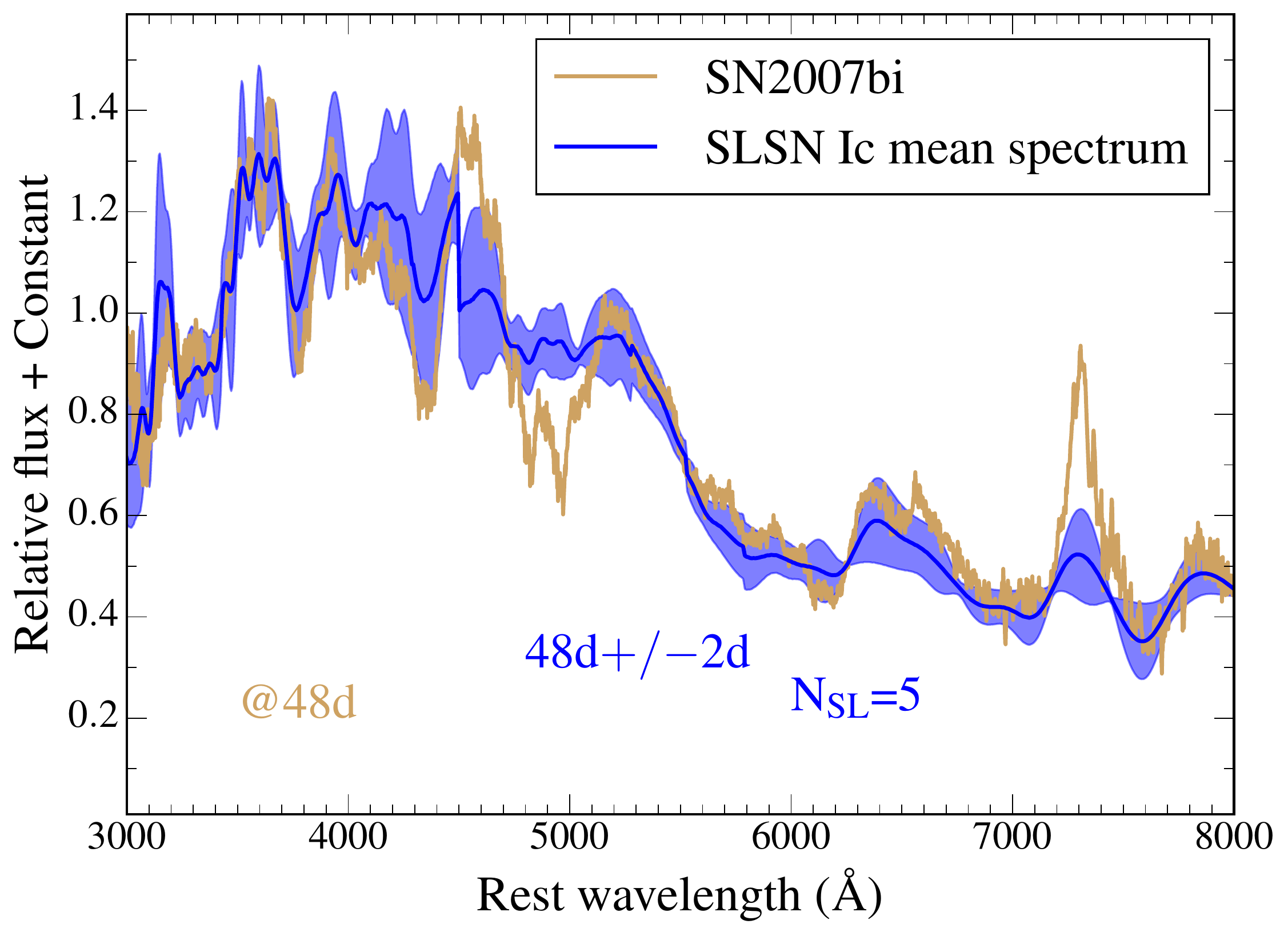}
\caption{Same as Figure \ref{mean_15lh}, but for spectrum of SN 2007bi at $t_{\mathrm{max}}=48$ days, as well as mean spectrum of SLSNe Ic at $t_{\mathrm{max}}=48\pm2$ days.}
\label{mean_07bi}
\end{figure}

SN 2007bi has been studied extensively and several models have been proposed for its origin: pair-instability explosion \citep{gal-yam09}, core-collapse explosion \citep{moriya10}, and magnetar model \citep{dessart12_slsn}. We note that the first spectrum taken for SN 2007bi is at $t_{\mathrm{max}}=48$ days. Thus, the velocity diagnostics as discussed in Section \ref{comp_model} could not be applied in this case. Some works claimed that SN 2007bi spectra are similar to other SLSN Ic spectra \citep{nicholl13, nicholl16}, while some concluded that SN 2007bi has redder spectra than other SLSNe Ic \citep{mazzali16}. Here, we try to address the question of whether SN 2007bi is a typical SLSN Ic from the spectral perspective by comparing its spectrum at $t_{\mathrm{max}}=48$ days to the average spectrum of SLSNe Ic at $t_{\mathrm{max}}=48\pm2$ days in Figure \ref{mean_07bi}. It is obvious that the spectrum of SN 2007bi is within one standard deviation of the SLSN Ic mean spectrum at most wavelengths, except for wavelength ranges 4500--5100 \AA~and 7200--7600 \AA, which are identified as due to Fe II lines and [Ca II] lines \citep{gal-yam09}, respectively. In particular, the Fe II region in the spectrum of SN 2007bi forms a distinct `W' feature, which is very different from the broad `U' feature in the average spectrum of SLSNe Ic. Thus, future works on SN 2007bi should be aware of the narrow and strong Fe II feature in absorption at $\sim4$800 \AA, as well as the extremely strong [Ca II] feature in emission at $\sim7$400 \AA. 

\section{Conclusions and Next steps}
\label{summary}

In this paper we have conducted a comprehensive spectroscopic comparison between the members of the full spectroscopic SN Ic family: SNe Ic, SNe Ic-bl, including those connected with GRBs, and SLSNe Ic. We have collected all of the published spectra of SLSNe Ic, quantified the diversity of their spectral features, and produced average spectra by using the novel methods we developed in \citet{liu16} and \citet{modjaz15}. Using the data products of SNe Ic-bl and of SNe Ic from \citet{modjaz15}, we have compared the population properties of these three members of SN Ic family in a systematic way. 

We have quantified the absorption velocities and FWHM widths of \FeIIn, not only because the feature is a common feature in SLSNe Ic, SNe Ic-bl, as well as in SNe Ic, but also because the feature has been suggested to trace the photospheric velocity. We have observed that the average absorption velocities and FWHM widths of \FeII are similar in SLSNe Ic and SNe Ic-bl, with both being systematically higher and broader than those in SNe Ic at all phases, thus strengthening the connection between SLSNe Ic and SN-GRBs. In particular, at $t_{\mathrm{max}}=10$ days, the weighted average \FeII absorption velocities in SLSNe Ic and SNe Ic-bl are $-15,000+/-2,600$ \kms~and $-18,500+/-7,400$ \kms, respectively, while the same velocities in SNe Ic amount to $-8,000+/-1,400$ \kms~on average. At the same phase, the \FeII features in SLSNe Ic and SNe Ic-bl are systematically and quantifiably broader than those in SNe Ic. Moreover, similarities between SLSNe Ic and SNe Ic-bl have been observed in host galaxy properties \citep{Lunnan14, Angus16} and nebular phase velocities \citep{jerkstrand16}. Thus, although the continuum of the SLSN Ic spectra is much bluer than that of the SN Ic-bl spectra, the many spectral and environmental similarities between SLSNe Ic and SNe Ic-bl, including those connected with GRBs, indicate that the two subtypes may have similar explosion engines and progenitors, which could be partly explained by a multi-dimensional magnetar model with different energy injection rates \citep{Metzger15, Suzuki16}. 

Two popular models to explain the power source of SLSNe Ic are the interaction model (e.g., \citealt{Chevalier11, Ginzburg12, ofek14}) and the magnetar model (e.g., \citealt{kasen10, woosley10, inserra13, nicholl13, Metzger15, chen16, Suzuki16}). However, our observations of high \FeII absorption velocities ($>1$0,000 \kms~for all measurements of SLSNe Ic between $t_{\mathrm{max}}=10$ days and $t_{\mathrm{max}}=20$ days), and broad \FeII width (FWHM $\sim1$0,000 \kms~with respect to that in SNe Ic at $t_{\mathrm{max}}=10$ days) in SLSNe Ic challenge the simple interaction model. Moreover, we find no objects out of 13 SLSNe Ic (with measured absorption velocities over a time span of more than 10 days) that have an unambiguous velocity plateau, at least as traced by the \FeII line. This finding, in addition to our earlier stated observations that the spectra of SLSNe Ic have broad lines similar to those of SNe Ic-bl, is inconsistent with the predictions of the one-dimensional magnetar model, and more consistent with the expected predictions of the two-dimensional magnetar model, though there are some caveats, which we discussed in Section \ref{caveat}. Given that the pair-instability model for SLSNe Ic is generally ruled out, except for SN 2007bi \citep{gal-yam09}, we conclude that none of the suggested three popular mechanisms for SLSNe Ic, or at least their simplified one-dimensional predictions, can fully explain our findings about the spectra of the full population of SLSNe Ic. 

Thus, as the next step, it would be necessary to compute the velocity structure from multi-dimensional magnetar models that have been developed for SLSNe Ic \citep{chen16, Suzuki16}. In addition, future synthesized spectra based on magnetar models need to reproduce the large observed line widths in SLSNe Ic as a population (see Table \ref{table_mean} and Figure \ref{mean_spec}), since current magnetar models predict as well as show only narrow or moderate line widths that are in line with those in SNe Ic and SNe Ib \citep{kasen10, dessart12}, but not with the majority of SLSNe Ic. In general, we look forward to modeling attempts that combine hydrodynamical modeling with non-LTE radiative transfer calculations, produce spectra as a function of time, and predict velocity evolution as well as other observables.

A narrow `w' feature around 4300 \AA~has been observed in SLSN Ic spectra at $t_\mathrm{max}<0$ day and identified as O II \citep{quimby07_05ap, quimby11, gal-yam12, nicholl15, mazzali16}. We have examined all spectra of SLSNe Ic, SNe Ic-bl, and SNe Ic before the date of maximum light, and found that only SLSN Ic spectra show the narrow `w' feature around 4300 \AA. Thus, the O II feature around 4300 \AA~provides a way to identify SLSNe Ic at early time (i.e., $t_{\mathrm{max}}<0$ day) as suggested by \citet{quimby11}. 

Finally, we have compared a few peculiar luminous objects to the SLSN Ic sample. We have compared the most luminous SN ASASSN-15lh to our average spectra, although we note that there is recent evidence against it being a SN (e.g., \citealt{leloudas16, Margutti16}). We have found that the spectra of ASASSN-15lh do not resemble average SLSN Ic spectra in terms of spectral features and continuum. We have also compared the SN spectral component of the first SN-ULGRB connection SN 2011kl / GRB 111209A with the average SN spectrum of SN-GRBs (i.e., SNe accompanying long-duration GRBs) and the average SLSNe Ic spectrum. We have found that the spectrum of SN 2011kl is consistent with the average spectrum of SLSNe Ic, but is much bluer than the average SN spectrum of SN-GRBs. In \citet{Greiner15}, a magnetar model is proposed to explain SN 2011kl. However, since only one spectrum of SN 2011kl was obtained, the velocity evolution is unknown and could not be used to check for any potential flat velocity-evolution as predicted in the one-dimensional magnetar model. Thus, before concluding that a magnetar model is responsible for SN 2011kl or SLSNe Ic in SN-ULGRB connections, more spectra are needed at multiple phases to verify the predicted velocity evolution based on the magnetar model.

One of the main outstanding questions in this field is whether there is a continuum in luminosity between SNe Ic and SLSNe Ic and how the luminosity distribution of SNe Ic-bl fits in this context, given the spectral connection between them and speculations by \citet{gal-yam12} that SNe Ic-bl may be intermediate events between radioactive powered SNe and engine powered SLSNe Ic. We suggest that future work tackles this question in a statistically robust way and includes SNe Ic-bl and SNe Ic from the same un-targeted surveys as the one that discovered SLSNe Ic.

All of the average spectra of SLSNe Ic and the weighted average velocities of \FeII at all phases presented in this manuscript can be downloaded from our group GitHub repository. \footnote{https://github.com/nyusngroup}

\acknowledgments
We are grateful to Andrew MacFadyen, Avishay Gal-Yam, Andrew Drake, Brian Metzger, Cosimo Inserra, Dan Kasen, Giorgos Leloudas, Matt Nicholl, Noam Soker, Stan Howerton, Stephan Hachinger, and Stan Woosley for useful discussions and suggestions.

Y.-Q. Liu is supported in part by NYU GSAS Dean's Dissertation Fellowship. M. Modjaz and the SNYU group are supported by the NSF CAREER award AST-1352405 and by the NSF award AST-1413260. 

This research has made use of NASA's Astrophysics Data System Bibliographic Services (ADS), the HyperLEDA database and the NASA/IPAC Extragalactic Database (NED) that is operated by the Jet Propulsion Laboratory, California Institute of Technology, under contract with the National Aeronautics and Space Administration. This paper has made extensive use of and the Weizmann interactive supernova data repository (www.weizmann.ac.il/astrophysics/wiserep)

\bibliographystyle{apj}
\bibliography{refs}

\begin{thebibliography}{}
\expandafter\ifx\csname natexlab\endcsname\relax\def\natexlab#1{#1}\fi

\bibitem[{{Angus} {et~al.}(2016){Angus}, {Levan}, {Perley}, {Tanvir}, {Lyman},
  {Stanway}, \& {Fruchter}}]{Angus16}
{Angus}, C.~R., {Levan}, A.~J., {Perley}, D.~A., {et~al.} 2016, \mnras, 458, 84

\bibitem[{{Arcavi} {et~al.}(2014){Arcavi}, {Gal-Yam}, {Sullivan}, {Pan},
  {Cenko}, {Horesh}, {Ofek}, {De Cia}, {Yan}, {Yang}, {Howell}, {Tal},
  {Kulkarni}, {Tendulkar}, {Tang}, {Xu}, {Sternberg}, {Cohen}, {Bloom},
  {Nugent}, {Kasliwal}, {Perley}, {Quimby}, {Miller}, {Theissen}, \&
  {Laher}}]{arcavi14}
{Arcavi}, I., {Gal-Yam}, A., {Sullivan}, M., {et~al.} 2014, \apj, 793, 38

\bibitem[{{Arcavi} {et~al.}(2016){Arcavi}, {Wolf}, {Howell}, {Bildsten},
  {Leloudas}, {Hardin}, {Prajs}, {Perley}, {Svirski}, {Gal-Yam}, {Katz},
  {McCully}, {Cenko}, {Lidman}, {Sullivan}, {Valenti}, {Astier}, {Balland},
  {Carlberg}, {Conley}, {Fouchez}, {Guy}, {Pain}, {Palanque-Delabrouille},
  {Perrett}, {Pritchet}, {Regnault}, {Rich}, \& {Ruhlmann-Kleider}}]{Arcavi16}
{Arcavi}, I., {Wolf}, W.~M., {Howell}, D.~A., {et~al.} 2016, \apj, 819, 35

\bibitem[{{Barbary} {et~al.}(2009){Barbary}, {Dawson}, {Tokita}, {Aldering},
  {Amanullah}, {Connolly}, {Doi}, {Faccioli}, {Fadeyev}, {Fruchter},
  {Goldhaber}, {Goobar}, {Gude}, {Huang}, {Ihara}, {Konishi}, {Kowalski},
  {Lidman}, {Meyers}, {Morokuma}, {Nugent}, {Perlmutter}, {Rubin}, {Schlegel},
  {Spadafora}, {Suzuki}, {Swift}, {Takanashi}, {Thomas}, \&
  {Yasuda}}]{Barbary09}
{Barbary}, K., {Dawson}, K.~S., {Tokita}, K., {et~al.} 2009, \apj, 690, 1358

\bibitem[{{Ben-Ami} {et~al.}(2014){Ben-Ami}, {Gal-Yam}, {Mazzali}, {Gnat},
  {Modjaz}, {Rabinak}, {Sullivan}, {Bildsten}, {Poznanski}, {Yaron}, {Arcavi},
  {Bloom}, {Horesh}, {Kasliwal}, {Kulkarni}, {Nugent}, {Ofek}, {Perley},
  {Quimby}, \& {Xu}}]{benami14}
{Ben-Ami}, S., {Gal-Yam}, A., {Mazzali}, P.~A., {et~al.} 2014, \apj, 785, 37

\bibitem[{{Bianco} {et~al.}(2014){Bianco}, {Modjaz}, {Hicken}, {Friedman},
  {Kirshner}, {Bloom}, {Challis}, {Marion}, {Wood-Vasey}, \& {Rest}}]{bianco14}
{Bianco}, F.~B., {Modjaz}, M., {Hicken}, M., {et~al.} 2014, \apjs, 213, 19

\bibitem[{{Branch} {et~al.}(2002){Branch}, {Benetti}, {Kasen}, {Baron},
  {Jeffery}, {Hatano}, {Stathakis}, {Filippenko}, {Matheson}, {Pastorello},
  {Altavilla}, {Cappellaro}, {Rizzi}, {Turatto}, {Li}, {Leonard}, \&
  {Shields}}]{branch02}
{Branch}, D., {Benetti}, S., {Kasen}, D., {et~al.} 2002, \apj, 566, 1005

\bibitem[{{Brown} {et~al.}(2016){Brown}, {Yang}, {Cooke}, {Olaes}, {Quimby},
  {Baade}, {Gehrels}, {Hoeflich}, {Maund}, {Mould}, {Patat}, {Wang}, \&
  {Wheeler}}]{brown16}
{Brown}, P.~J., {Yang}, Y., {Cooke}, J., {et~al.} 2016, ArXiv e-prints,
  arXiv:1605.03951

\bibitem[{{Cano}(2013)}]{cano13}
{Cano}, Z. 2013, \mnras, 434, 1098

\bibitem[{{Cano} {et~al.}(2016){Cano}, {Wang}, {Dai}, \& {Wu}}]{cano16}
{Cano}, Z., {Wang}, S.-Q., {Dai}, Z.-G., \& {Wu}, X.-F. 2016, ArXiv e-prints,
  arXiv:1604.03549

\bibitem[{{Chatzopoulos} {et~al.}(2013){Chatzopoulos}, {Wheeler}, {Vinko},
  {Horvath}, \& {Nagy}}]{Chatzopoulos13}
{Chatzopoulos}, E., {Wheeler}, J.~C., {Vinko}, J., {Horvath}, Z.~L., \& {Nagy},
  A. 2013, \apj, 773, 76

\bibitem[{{Chatzopoulos} {et~al.}(2016){Chatzopoulos}, {Wheeler}, {Vinko},
  {Nagy}, {Wiggins}, \& {Even}}]{Chatzopoulos16}
{Chatzopoulos}, E., {Wheeler}, J.~C., {Vinko}, J., {et~al.} 2016, \apj, 828, 94

\bibitem[{{Chen} {et~al.}(2016{\natexlab{a}}){Chen}, {Woosley}, \&
  {Sukhbold}}]{chen16}
{Chen}, K.-J., {Woosley}, S.~E., \& {Sukhbold}, T. 2016{\natexlab{a}}, \apj,
  832, 73

\bibitem[{{Chen} {et~al.}(2013){Chen}, {Smartt}, {Bresolin}, {Pastorello},
  {Kudritzki}, {Kotak}, {McCrum}, {Fraser}, \& {Valenti}}]{chen13}
{Chen}, T.-W., {Smartt}, S.~J., {Bresolin}, F., {et~al.} 2013, \apjl, 763, L28

\bibitem[{{Chen} {et~al.}(2016{\natexlab{b}}){Chen}, {Nicholl}, {Smartt},
  {Mazzali}, {Yates}, {Moriya}, {Inserra}, {Langer}, {Kruehler}, {Pan},
  {Kotak}, {Galbany}, {Schady}, {Wiseman}, {Greiner}, {Schulze}, {Man},
  {Jerkstrand}, {Smith}, {Dennefeld}, {Baltay}, {Bolmer}, {Kankare}, {Knust},
  {Maguire}, {Rabinowitz}, {Rostami}, {Sullivan}, \& {Young}}]{chen2016}
{Chen}, T.-W., {Nicholl}, M., {Smartt}, S.~J., {et~al.} 2016{\natexlab{b}},
  ArXiv e-prints, arXiv:1611.09910

\bibitem[{{Chevalier} \& {Irwin}(2011)}]{Chevalier11}
{Chevalier}, R.~A., \& {Irwin}, C.~M. 2011, \apjl, 729, L6

\bibitem[{{Chomiuk} {et~al.}(2011){Chomiuk}, {Chornock}, {Soderberg}, {Berger},
  {Chevalier}, {Foley}, {Huber}, {Narayan}, {Rest}, {Gezari}, {Kirshner},
  {Riess}, {Rodney}, {Smartt}, {Stubbs}, {Tonry}, {Wood-Vasey}, {Burgett},
  {Chambers}, {Czekala}, {Flewelling}, {Forster}, {Kaiser}, {Kudritzki},
  {Magnier}, {Martin}, {Morgan}, {Neill}, {Price}, {Roth}, {Sanders}, \&
  {Wainscoat}}]{chomiuk11}
{Chomiuk}, L., {Chornock}, R., {Soderberg}, A.~M., {et~al.} 2011, \apj, 743,
  114

\bibitem[{{Chornock} {et~al.}(2013){Chornock}, {Berger}, {Rest},
  {Milisavljevic}, {Lunnan}, {Foley}, {Soderberg}, {Smartt}, {Burgasser},
  {Challis}, {Chomiuk}, {Czekala}, {Drout}, {Fong}, {Huber}, {Kirshner},
  {Leibler}, {McLeod}, {Marion}, {Narayan}, {Riess}, {Roth}, {Sanders},
  {Scolnic}, {Smith}, {Stubbs}, {Tonry}, {Valenti}, {Burgett}, {Chambers},
  {Hodapp}, {Kaiser}, {Kudritzki}, {Magnier}, \& {Price}}]{chornock13}
{Chornock}, R., {Berger}, E., {Rest}, A., {et~al.} 2013, \apj, 767, 162

\bibitem[{{Dessart} {et~al.}(2012{\natexlab{a}}){Dessart}, {Hillier}, {Li}, \&
  {Woosley}}]{dessart12}
{Dessart}, L., {Hillier}, D.~J., {Li}, C., \& {Woosley}, S. 2012{\natexlab{a}},
  \mnras, 424, 2139

\bibitem[{{Dessart} {et~al.}(2012{\natexlab{b}}){Dessart}, {Hillier},
  {Waldman}, {Livne}, \& {Blondin}}]{dessart12_slsn}
{Dessart}, L., {Hillier}, D.~J., {Waldman}, R., {Livne}, E., \& {Blondin}, S.
  2012{\natexlab{b}}, \mnras, 426, L76

\bibitem[{{Dessart} {et~al.}(2015){Dessart}, {Hillier}, {Woosley}, {Livne},
  {Waldman}, {Yoon}, \& {Langer}}]{dessart15}
{Dessart}, L., {Hillier}, D.~J., {Woosley}, S., {et~al.} 2015, ArXiv e-prints,
  arXiv:1507.07783

\bibitem[{{Dessart} {et~al.}(2016){Dessart}, {Hillier}, {Woosley}, {Livne},
  {Waldman}, {Yoon}, \& {Langer}}]{dessart16}
---. 2016, \mnras, 458, 1618

\bibitem[{{Dessart} {et~al.}(2013){Dessart}, {Waldman}, {Livne}, {Hillier}, \&
  {Blondin}}]{dessart13}
{Dessart}, L., {Waldman}, R., {Livne}, E., {Hillier}, D.~J., \& {Blondin}, S.
  2013, \mnras, 428, 3227

\bibitem[{{Dexter} \& {Kasen}(2013)}]{Dexter13}
{Dexter}, J., \& {Kasen}, D. 2013, \apj, 772, 30

\bibitem[{{Dong} {et~al.}(2016){Dong}, {Shappee}, {Prieto}, {Jha}, {Stanek},
  {Holoien}, {Kochanek}, {Thompson}, {Morrell}, {Thompson}, {Basu}, {Beacom},
  {Bersier}, {Brimacombe}, {Brown}, {Bufano}, {Chen}, {Conseil}, {Danilet},
  {Falco}, {Grupe}, {Kiyota}, {Masi}, {Nicholls}, {Olivares E.}, {Pignata},
  {Pojmanski}, {Simonian}, {Szczygiel}, \& {Wo{\'z}niak}}]{dong16}
{Dong}, S., {Shappee}, B.~J., {Prieto}, J.~L., {et~al.} 2016, Science, 351, 257

\bibitem[{{Drout} {et~al.}(2011){Drout}, {Soderberg}, {Gal-Yam}, {Cenko},
  {Fox}, {Leonard}, {Sand}, {Moon}, {Arcavi}, \& {Green}}]{drout11}
{Drout}, M.~R., {Soderberg}, A.~M., {Gal-Yam}, A., {et~al.} 2011, \apj, 741, 97

\bibitem[{{Folatelli} {et~al.}(2014){Folatelli}, {Bersten}, {Kuncarayakti},
  {Olivares Estay}, {Anderson}, {Holmbo}, {Maeda}, {Morrell}, {Nomoto},
  {Pignata}, {Stritzinger}, {Contreras}, {F{\"o}rster}, {Hamuy}, {Phillips},
  {Prieto}, {Valenti}, {Afonso}, {Altenm{\"u}ller}, {Elliott}, {Greiner},
  {Updike}, {Haislip}, {LaCluyze}, {Moore}, \& {Reichart}}]{folatelli14}
{Folatelli}, G., {Bersten}, M.~C., {Kuncarayakti}, H., {et~al.} 2014, \apj,
  792, 7

\bibitem[{{Foley} {et~al.}(2007){Foley}, {Smith}, {Ganeshalingam}, {Li},
  {Chornock}, \& {Filippenko}}]{foley07}
{Foley}, R.~J., {Smith}, N., {Ganeshalingam}, M., {et~al.} 2007, \apjl, 657,
  L105

\bibitem[{{Gal-Yam}(2012)}]{gal-yam12}
{Gal-Yam}, A. 2012, Science, 337, 927

\bibitem[{{Gal-Yam}(2016)}]{gal-yam16}
---. 2016, ArXiv e-prints, arXiv:1611.09353

\bibitem[{{Gal-Yam} {et~al.}(2009){Gal-Yam}, {Mazzali}, {Ofek}, {Nugent},
  {Kulkarni}, {Kasliwal}, {Quimby}, {Filippenko}, {Cenko}, {Chornock},
  {Waldman}, {Kasen}, {Sullivan}, {Beshore}, {Drake}, {Thomas}, {Bloom},
  {Poznanski}, {Miller}, {Foley}, {Silverman}, {Arcavi}, {Ellis}, \&
  {Deng}}]{gal-yam09}
{Gal-Yam}, A., {Mazzali}, P., {Ofek}, E.~O., {et~al.} 2009, \nat, 462, 624

\bibitem[{{Gilkis} {et~al.}(2016){Gilkis}, {Soker}, \& {Papish}}]{Gilkis16}
{Gilkis}, A., {Soker}, N., \& {Papish}, O. 2016, \apj, 826, 178

\bibitem[{{Ginzburg} \& {Balberg}(2012)}]{Ginzburg12}
{Ginzburg}, S., \& {Balberg}, S. 2012, \apj, 757, 178

\bibitem[{{Godoy-Rivera} {et~al.}(2016){Godoy-Rivera}, {Stanek}, {Kochanek},
  {Chen}, {Dong}, {Prieto}, {Shappee}, {Jha}, {Foley}, {Pan}, {Holoien},
  {Thompson}, {Grupe}, \& {Beacom}}]{Godoy-Rivera16}
{Godoy-Rivera}, D., {Stanek}, K.~Z., {Kochanek}, C.~S., {et~al.} 2016, ArXiv
  e-prints, arXiv:1605.00645

\bibitem[{{Greiner} {et~al.}(2015){Greiner}, {Mazzali}, {Kann}, {Kr{\"u}hler},
  {Pian}, {Prentice}, {Olivares E.}, {Rossi}, {Klose}, {Taubenberger}, {Knust},
  {Afonso}, {Ashall}, {Bolmer}, {Delvaux}, {Diehl}, {Elliott}, {Filgas},
  {Fynbo}, {Graham}, {Guelbenzu}, {Kobayashi}, {Leloudas}, {Savaglio},
  {Schady}, {Schmidl}, {Schweyer}, {Sudilovsky}, {Tanga}, {Updike}, {van
  Eerten}, \& {Varela}}]{Greiner15}
{Greiner}, J., {Mazzali}, P.~A., {Kann}, D.~A., {et~al.} 2015, \nat, 523, 189

\bibitem[{{Hachinger} {et~al.}(2012){Hachinger}, {Mazzali}, {Taubenberger},
  {Fink}, {Pakmor}, {Hillebrandt}, \& {Seitenzahl}}]{hachinger12_Ia}
{Hachinger}, S., {Mazzali}, P.~A., {Taubenberger}, S., {et~al.} 2012, \mnras,
  427, 2057

\bibitem[{{Hjorth} \& {Bloom}(2012)}]{Hjorth12}
{Hjorth}, J., \& {Bloom}, J.~S. 2012, {The Gamma-Ray Burst - Supernova
  Connection}, 169--190

\bibitem[{{Howell} {et~al.}(2013){Howell}, {Kasen}, {Lidman}, {Sullivan},
  {Conley}, {Astier}, {Balland}, {Carlberg}, {Fouchez}, {Guy}, {Hardin},
  {Pain}, {Palanque-Delabrouille}, {Perrett}, {Pritchet}, {Regnault}, {Rich},
  \& {Ruhlmann-Kleider}}]{howell13}
{Howell}, D.~A., {Kasen}, D., {Lidman}, C., {et~al.} 2013, \apj, 779, 98

\bibitem[{{Inserra} {et~al.}(2016){Inserra}, {Bulla}, {Sim}, \&
  {Smartt}}]{Inserra16}
{Inserra}, C., {Bulla}, M., {Sim}, S.~A., \& {Smartt}, S.~J. 2016, \apj, 831,
  79

\bibitem[{{Inserra} {et~al.}(2013){Inserra}, {Smartt}, {Jerkstrand}, {Valenti},
  {Fraser}, {Wright}, {Smith}, {Chen}, {Kotak}, {Pastorello}, {Nicholl},
  {Bresolin}, {Kudritzki}, {Benetti}, {Botticella}, {Burgett}, {Chambers},
  {Ergon}, {Flewelling}, {Fynbo}, {Geier}, {Hodapp}, {Howell}, {Huber},
  {Kaiser}, {Leloudas}, {Magill}, {Magnier}, {McCrum}, {Metcalfe}, {Price},
  {Rest}, {Sollerman}, {Sweeney}, {Taddia}, {Taubenberger}, {Tonry},
  {Wainscoat}, {Waters}, \& {Young}}]{inserra13}
{Inserra}, C., {Smartt}, S.~J., {Jerkstrand}, A., {et~al.} 2013, \apj, 770, 128

\bibitem[{{Inserra} {et~al.}(2017){Inserra}, {Nicholl}, {Chen}, {Jerkstrand},
  {Smartt}, {Kr{\"u}hler}, {Anderson}, {Baltay}, {Della Valle}, {Fraser},
  {Gal-Yam}, {Galbany}, {Kankare}, {Maguire}, {Rabinowitz}, {Smith}, {Valenti},
  \& {Young}}]{Inserra17}
{Inserra}, C., {Nicholl}, M., {Chen}, T.-W., {et~al.} 2017, ArXiv e-prints,
  arXiv:1701.00941

\bibitem[{{Jerkstrand} {et~al.}(2016){Jerkstrand}, {Smartt}, {Inserra},
  {Nicholl}, {Chen}, {Kr{\"u}hler}, {Sollerman}, {Taubenberger}, {Gal-Yam},
  {Kankare}, {Maguire}, {Fraser}, {Valenti}, {Sullivan}, {Cartier}, \&
  {Young}}]{jerkstrand16}
{Jerkstrand}, A., {Smartt}, S.~J., {Inserra}, C., {et~al.} 2016, ArXiv
  e-prints, arXiv:1608.02994

\bibitem[{{Kann} {et~al.}(2016){Kann}, {Schady}, {Olivares E.}, {Klose},
  {Rossi}, {Perley}, {Kr{\"u}hler}, {Greiner}, {Nicuesa Guelbenzu}, {Elliott},
  {Knust}, {Filgas}, {Pian}, {Mazzali}, {Fynbo}, {Leloudas}, {Afonso},
  {Delvaux}, {Graham}, {Rau}, {Schmidl}, {Schulze}, {Tanga}, {Updike}, \&
  {Varela}}]{Kann16}
{Kann}, D.~A., {Schady}, P., {Olivares E.}, F., {et~al.} 2016, ArXiv e-prints,
  arXiv:1606.06791

\bibitem[{{Kasen} \& {Bildsten}(2010)}]{kasen10}
{Kasen}, D., \& {Bildsten}, L. 2010, \apj, 717, 245

\bibitem[{{Kelly} \& {Kirshner}(2012)}]{kelly12}
{Kelly}, P.~L., \& {Kirshner}, R.~P. 2012, \apj, 759, 107

\bibitem[{{Leloudas} {et~al.}(2012){Leloudas}, {Chatzopoulos}, {Dilday},
  {Gorosabel}, {Vinko}, {Gallazzi}, {Wheeler}, {Bassett}, {Fischer}, {Frieman},
  {Fynbo}, {Goobar}, {Jel{\'{\i}}nek}, {Malesani}, {Nichol}, {Nordin},
  {{\"O}stman}, {Sako}, {Schneider}, {Smith}, {Sollerman}, {Stritzinger},
  {Th{\"o}ne}, \& {de Ugarte Postigo}}]{leloudas12}
{Leloudas}, G., {Chatzopoulos}, E., {Dilday}, B., {et~al.} 2012, \aap, 541,
  A129

\bibitem[{{Leloudas} {et~al.}(2015){Leloudas}, {Schulze}, {Kr{\"u}hler},
  {Gorosabel}, {Christensen}, {Mehner}, {de Ugarte Postigo}, {Amor{\'{\i}}n},
  {Th{\"o}ne}, {Anderson}, {Bauer}, {Gallazzi}, {He{\l}miniak}, {Hjorth},
  {Ibar}, {Malesani}, {Morell}, {Vinko}, \& {Wheeler}}]{Leloudas15}
{Leloudas}, G., {Schulze}, S., {Kr{\"u}hler}, T., {et~al.} 2015, \mnras, 449,
  917

\bibitem[{{Leloudas} {et~al.}(2016){Leloudas}, {Fraser}, {Stone}, {van Velzen},
  {Jonker}, {Arcavi}, {Fremling}, {Maund}, {Smartt}, {Kruhler}, {Miller-Jones},
  {Vreeswijk}, {Gal-Yam}, {Mazzali}, {De Cia}, {Howell}, {Inserra}, {Patat},
  {de Ugarte Postigo}, {Yaron}, {Ashall}, {Bar}, {Campbell}, {Chen},
  {Childress}, {Elias-Rosa}, {Harmanen}, {Hosseinzadeh}, {Johansson}, {Kangas},
  {Kankare}, {Kim}, {Kuncarayakti}, {Lyman}, {Magee}, {Maguire}, {Malesani},
  {Mattila}, {McCully}, {Nicholl}, {Prentice}, {Romero-Canizales}, {Schulze},
  {Smith}, {Sollerman}, {Sullivan}, {Tucker}, {Valenti}, {Wheeler}, \&
  {Young}}]{leloudas16}
{Leloudas}, G., {Fraser}, M., {Stone}, N.~C., {et~al.} 2016, ArXiv e-prints,
  arXiv:1609.02927

\bibitem[{{Leloudas} {et~al.}(2017){Leloudas}, {Maund}, {Gal-Yam}, {Pursimo},
  {Hsiao}, {Malesani}, {Patat}, {de Ugarte Postigo}, {Sollerman},
  {Stritzinger}, \& {Wheeler}}]{Leloudas17}
{Leloudas}, G., {Maund}, J.~R., {Gal-Yam}, A., {et~al.} 2017, \apjl, 837, L14

\bibitem[{{Liu} {et~al.}(2016){Liu}, {Modjaz}, {Bianco}, \& {Graur}}]{liu16}
{Liu}, Y.-Q., {Modjaz}, M., {Bianco}, F.~B., \& {Graur}, O. 2016, \apj, 827, 90

\bibitem[{{Lunnan} {et~al.}(2013){Lunnan}, {Chornock}, {Berger},
  {Milisavljevic}, {Drout}, {Sanders}, {Challis}, {Czekala}, {Foley}, {Fong},
  {Huber}, {Kirshner}, {Leibler}, {Marion}, {McCrum}, {Narayan}, {Rest},
  {Roth}, {Scolnic}, {Smartt}, {Smith}, {Soderberg}, {Stubbs}, {Tonry},
  {Burgett}, {Chambers}, {Kudritzki}, {Magnier}, \& {Price}}]{lunnan13}
{Lunnan}, R., {Chornock}, R., {Berger}, E., {et~al.} 2013, \apj, 771, 97

\bibitem[{{Lunnan} {et~al.}(2014){Lunnan}, {Chornock}, {Berger}, {Laskar},
  {Fong}, {Rest}, {Sanders}, {Challis}, {Drout}, {Foley}, {Huber}, {Kirshner},
  {Leibler}, {Marion}, {McCrum}, {Milisavljevic}, {Narayan}, {Scolnic},
  {Smartt}, {Smith}, {Soderberg}, {Tonry}, {Burgett}, {Chambers}, {Flewelling},
  {Hodapp}, {Kaiser}, {Magnier}, {Price}, \& {Wainscoat}}]{Lunnan14}
---. 2014, \apj, 787, 138

\bibitem[{{Lyman} {et~al.}(2016){Lyman}, {Bersier}, {James}, {Mazzali},
  {Eldridge}, {Fraser}, \& {Pian}}]{lyman14}
{Lyman}, J.~D., {Bersier}, D., {James}, P.~A., {et~al.} 2016, \mnras, 457, 328

\bibitem[{{Maier} {et~al.}(2015){Maier}, {Ziegler}, {Lilly}, {Contini},
  {P{\'e}rez-Montero}, {Lamareille}, {Bolzonella}, \& {Le Floc'h}}]{Maier15}
{Maier}, C., {Ziegler}, B.~L., {Lilly}, S.~J., {et~al.} 2015, \aap, 577, A14

\bibitem[{{Margutti} {et~al.}(2016){Margutti}, {Metzger}, {Chornock},
  {Milisavljevic}, {Berger}, {Blanchard}, {Guidorzi}, {Migliori}, {Kamble},
  {Lunnan}, {Nicholl}, {Coppejans}, {Dall'Osso}, {Drout}, {Perna}, \&
  {Sbarufatti}}]{Margutti16}
{Margutti}, R., {Metzger}, B.~D., {Chornock}, R., {et~al.} 2016, ArXiv
  e-prints, arXiv:1610.01632

\bibitem[{{Mazzali} {et~al.}(2016){Mazzali}, {Sullivan}, {Pian}, {Greiner}, \&
  {Kann}}]{mazzali16}
{Mazzali}, P.~A., {Sullivan}, M., {Pian}, E., {Greiner}, J., \& {Kann}, D.~A.
  2016, \mnras, 458, 3455

\bibitem[{{McCrum} {et~al.}(2014){McCrum}, {Smartt}, {Kotak}, {Rest},
  {Jerkstrand}, {Inserra}, {Rodney}, {Chen}, {Howell}, {Huber}, {Pastorello},
  {Tonry}, {Bresolin}, {Kudritzki}, {Chornock}, {Berger}, {Smith},
  {Botticella}, {Foley}, {Fraser}, {Milisavljevic}, {Nicholl}, {Riess},
  {Stubbs}, {Valenti}, {Wood-Vasey}, {Wright}, {Young}, {Drout}, {Czekala},
  {Burgett}, {Chambers}, {Draper}, {Flewelling}, {Hodapp}, {Kaiser}, {Magnier},
  {Metcalfe}, {Price}, {Sweeney}, \& {Wainscoat}}]{McCrum14}
{McCrum}, M., {Smartt}, S.~J., {Kotak}, R., {et~al.} 2014, \mnras, 437, 656

\bibitem[{{Metzger} {et~al.}(2015){Metzger}, {Margalit}, {Kasen}, \&
  {Quataert}}]{Metzger15}
{Metzger}, B.~D., {Margalit}, B., {Kasen}, D., \& {Quataert}, E. 2015, \mnras,
  454, 3311

\bibitem[{{Modjaz}(2012)}]{modjaz12_proc}
{Modjaz}, M. 2012, in IAU Symposium, Vol. 279, IAU Symposium, 207--211

\bibitem[{{Modjaz} {et~al.}(2011){Modjaz}, {Kewley}, {Bloom}, {Filippenko},
  {Perley}, \& {Silverman}}]{modjaz11}
{Modjaz}, M., {Kewley}, L., {Bloom}, J.~S., {et~al.} 2011, \apjl, 731, L4

\bibitem[{{Modjaz} {et~al.}(2016){Modjaz}, {Liu}, {Bianco}, \&
  {Graur}}]{modjaz15}
{Modjaz}, M., {Liu}, Y.~Q., {Bianco}, F.~B., \& {Graur}, O. 2016, \apj, 832,
  108

\bibitem[{{Modjaz} {et~al.}(2008){Modjaz}, {Kewley}, {Kirshner}, {Stanek},
  {Challis}, {Garnavich}, {Greene}, {Kelly}, \& {Prieto}}]{modjaz08_Z}
{Modjaz}, M., {Kewley}, L., {Kirshner}, R.~P., {et~al.} 2008, \aj, 135, 1136

\bibitem[{{Modjaz} {et~al.}(2009){Modjaz}, {Li}, {Butler}, {Chornock},
  {Perley}, {Blondin}, {Bloom}, {Filippenko}, {Kirshner}, {Kocevski},
  {Poznanski}, {Hicken}, {Foley}, {Stringfellow}, {Berlind}, {Barrado y
  Navascues}, {Blake}, {Bouy}, {Brown}, {Challis}, {Chen}, {de Vries},
  {Dufour}, {Falco}, {Friedman}, {Ganeshalingam}, {Garnavich}, {Holden},
  {Illingworth}, {Lee}, {Liebert}, {Marion}, {Olivier}, {Prochaska},
  {Silverman}, {Smith}, {Starr}, {Steele}, {Stockton}, {Williams}, \&
  {Wood-Vasey}}]{modjaz09}
{Modjaz}, M., {Li}, W., {Butler}, N., {et~al.} 2009, \apj, 702, 226

\bibitem[{{Moriya} {et~al.}(2010){Moriya}, {Tominaga}, {Tanaka}, {Maeda}, \&
  {Nomoto}}]{moriya10}
{Moriya}, T., {Tominaga}, N., {Tanaka}, M., {Maeda}, K., \& {Nomoto}, K. 2010,
  \apjl, 717, L83

\bibitem[{{Nicholl} {et~al.}(2013){Nicholl}, {Smartt}, {Jerkstrand}, {Inserra},
  {McCrum}, {Kotak}, {Fraser}, {Wright}, {Chen}, {Smith}, {Young}, {Sim},
  {Valenti}, {Howell}, {Bresolin}, {Kudritzki}, {Tonry}, {Huber}, {Rest},
  {Pastorello}, {Tomasella}, {Cappellaro}, {Benetti}, {Mattila}, {Kankare},
  {Kangas}, {Leloudas}, {Sollerman}, {Taddia}, {Berger}, {Chornock}, {Narayan},
  {Stubbs}, {Foley}, {Lunnan}, {Soderberg}, {Sanders}, {Milisavljevic},
  {Margutti}, {Kirshner}, {Elias-Rosa}, {Morales-Garoffolo}, {Taubenberger},
  {Botticella}, {Gezari}, {Urata}, {Rodney}, {Riess}, {Scolnic}, {Wood-Vasey},
  {Burgett}, {Chambers}, {Flewelling}, {Magnier}, {Kaiser}, {Metcalfe},
  {Morgan}, {Price}, {Sweeney}, \& {Waters}}]{nicholl13}
{Nicholl}, M., {Smartt}, S.~J., {Jerkstrand}, A., {et~al.} 2013, \nat, 502, 346

\bibitem[{{Nicholl} {et~al.}(2014){Nicholl}, {Smartt}, {Jerkstrand}, {Inserra},
  {Anderson}, {Baltay}, {Benetti}, {Chen}, {Elias-Rosa}, {Feindt}, {Fraser},
  {Gal-Yam}, {Hadjiyska}, {Howell}, {Kotak}, {Lawrence}, {Leloudas},
  {Margheim}, {Mattila}, {McCrum}, {McKinnon}, {Mead}, {Nugent}, {Rabinowitz},
  {Rest}, {Smith}, {Sollerman}, {Sullivan}, {Taddia}, {Valenti}, {Walker}, \&
  {Young}}]{nicholl14}
---. 2014, \mnras, 444, 2096

\bibitem[{{Nicholl} {et~al.}(2015{\natexlab{a}}){Nicholl}, {Smartt},
  {Jerkstrand}, {Sim}, {Inserra}, {Anderson}, {Baltay}, {Benetti}, {Chambers},
  {Chen}, {Elias-Rosa}, {Feindt}, {Flewelling}, {Fraser}, {Gal-Yam}, {Galbany},
  {Huber}, {Kangas}, {Kankare}, {Kotak}, {Kr{\"u}hler}, {Maguire}, {McKinnon},
  {Rabinowitz}, {Rostami}, {Schulze}, {Smith}, {Sullivan}, {Tonry}, {Valenti},
  \& {Young}}]{nicholl15_lsq14bdq}
---. 2015{\natexlab{a}}, \apjl, 807, L18

\bibitem[{{Nicholl} {et~al.}(2015{\natexlab{b}}){Nicholl}, {Smartt},
  {Jerkstrand}, {Inserra}, {Sim}, {Chen}, {Benetti}, {Fraser}, {Gal-Yam},
  {Kankare}, {Maguire}, {Smith}, {Sullivan}, {Valenti}, {Young}, {Baltay},
  {Bauer}, {Baumont}, {Bersier}, {Botticella}, {Childress}, {Dennefeld}, {Della
  Valle}, {Elias-Rosa}, {Feindt}, {Galbany}, {Hadjiyska}, {Le Guillou},
  {Leloudas}, {Mazzali}, {McKinnon}, {Polshaw}, {Rabinowitz}, {Rostami},
  {Scalzo}, {Schmidt}, {Schulze}, {Sollerman}, {Taddia}, \& {Yuan}}]{nicholl15}
---. 2015{\natexlab{b}}, \mnras, 452, 3869

\bibitem[{{Nicholl} {et~al.}(2016){Nicholl}, {Berger}, {Smartt}, {Margutti},
  {Kamble}, {Alexander}, {Chen}, {Inserra}, {Arcavi}, {Blanchard}, {Cartier},
  {Chambers}, {Childress}, {Chornock}, {Cowperthwaite}, {Drout}, {Flewelling},
  {Fraser}, {Gal-Yam}, {Galbany}, {Harmanen}, {Holoien}, {Hosseinzadeh},
  {Howell}, {Huber}, {Jerkstrand}, {Kankare}, {Kochanek}, {Lin}, {Lunnan},
  {Magnier}, {Maguire}, {McCully}, {McDonald}, {Metzger}, {Milisavljevic},
  {Mitra}, {Reynolds}, {Saario}, {Shappee}, {Smith}, {Valenti}, {Villar},
  {Waters}, \& {Young}}]{nicholl16}
{Nicholl}, M., {Berger}, E., {Smartt}, S.~J., {et~al.} 2016, ArXiv e-prints,
  arXiv:1603.04748

\bibitem[{{Ofek} {et~al.}(2014){Ofek}, {Arcavi}, {Tal}, {Sullivan}, {Gal-Yam},
  {Kulkarni}, {Nugent}, {Ben-Ami}, {Bersier}, {Cao}, {Cenko}, {De Cia},
  {Filippenko}, {Fransson}, {Kasliwal}, {Laher}, {Surace}, {Quimby}, \&
  {Yaron}}]{ofek14}
{Ofek}, E.~O., {Arcavi}, I., {Tal}, D., {et~al.} 2014, \apj, 788, 154

\bibitem[{{Omand} {et~al.}(2017){Omand}, {Kashiyama}, \& {Murase}}]{Omand2017}
{Omand}, C.~M.~B., {Kashiyama}, K., \& {Murase}, K. 2017, ArXiv e-prints,
  arXiv:1704.00456

\bibitem[{{Papadopoulos} {et~al.}(2015){Papadopoulos}, {D'Andrea}, {Sullivan},
  {Nichol}, {Barbary}, {Biswas}, {Brown}, {Covarrubias}, {Finley}, {Fischer},
  {Foley}, {Goldstein}, {Gupta}, {Kessler}, {Kovacs}, {Kuhlmann}, {Lidman},
  {March}, {Nugent}, {Sako}, {Smith}, {Spinka}, {Wester}, {Abbott}, {Abdalla},
  {Allam}, {Banerji}, {Bernstein}, {Bernstein}, {Carnero}, {da Costa}, {DePoy},
  {Desai}, {Diehl}, {Eifler}, {Evrard}, {Flaugher}, {Frieman}, {Gerdes},
  {Gruen}, {Honscheid}, {James}, {Kuehn}, {Kuropatkin}, {Lahav}, {Maia},
  {Makler}, {Marshall}, {Merritt}, {Miller}, {Miquel}, {Ogando}, {Plazas},
  {Roe}, {Romer}, {Rykoff}, {Sanchez}, {Santiago}, {Scarpine}, {Schubnell},
  {Sevilla}, {Soares-Santos}, {Suchyta}, {Swanson}, {Tarle}, {Thaler},
  {Tucker}, {Wechsler}, \& {Zuntz}}]{Papadopoulos15}
{Papadopoulos}, A., {D'Andrea}, C.~B., {Sullivan}, M., {et~al.} 2015, \mnras,
  449, 1215

\bibitem[{{Pastorello} {et~al.}(2008){Pastorello}, {Quimby}, {Smartt},
  {Mattila}, {Navasardyan}, {Crockett}, {Elias-Rosa}, {Mondol}, {Wheeler}, \&
  {Young}}]{pastorello08}
{Pastorello}, A., {Quimby}, R.~M., {Smartt}, S.~J., {et~al.} 2008, \mnras, 389,
  131

\bibitem[{{Pastorello} {et~al.}(2010){Pastorello}, {Smartt}, {Botticella},
  {Maguire}, {Fraser}, {Smith}, {Kotak}, {Magill}, {Valenti}, {Young},
  {Gezari}, {Bresolin}, {Kudritzki}, {Howell}, {Rest}, {Metcalfe}, {Mattila},
  {Kankare}, {Huang}, {Urata}, {Burgett}, {Chambers}, {Dombeck}, {Flewelling},
  {Grav}, {Heasley}, {Hodapp}, {Kaiser}, {Luppino}, {Lupton}, {Magnier},
  {Monet}, {Morgan}, {Onaka}, {Price}, {Rhoads}, {Siegmund}, {Stubbs},
  {Sweeney}, {Tonry}, {Wainscoat}, {Waterson}, {Waters}, \&
  {Wynn-Williams}}]{Pastorello10}
{Pastorello}, A., {Smartt}, S.~J., {Botticella}, M.~T., {et~al.} 2010, \apjl,
  724, L16

\bibitem[{{Pastorello} {et~al.}(2016){Pastorello}, {Wang}, {Ciabattari},
  {Bersier}, {Mazzali}, {Gao}, {Xu}, {Zhang}, {Tokuoka}, {Benetti},
  {Cappellaro}, {Elias-Rosa}, {Harutyunyan}, {Huang}, {Miluzio}, {Mo},
  {Ochner}, {Tartaglia}, {Terreran}, {Tomasella}, \& {Turatto}}]{pastorello16}
{Pastorello}, A., {Wang}, X.-F., {Ciabattari}, F., {et~al.} 2016, \mnras, 456,
  853

\bibitem[{{Perley} {et~al.}(2016){Perley}, {Quimby}, {Yan}, {Vreeswijk}, {De
  Cia}, {Lunnan}, {Gal-Yam}, {Yaron}, {Filippenko}, {Graham}, {Laher}, \&
  {Nugent}}]{perley16}
{Perley}, D.~A., {Quimby}, R.~M., {Yan}, L., {et~al.} 2016, \apj, 830, 13

\bibitem[{{Quimby} {et~al.}(2007){Quimby}, {Aldering}, {Wheeler},
  {H{\"o}flich}, {Akerlof}, \& {Rykoff}}]{quimby07_05ap}
{Quimby}, R.~M., {Aldering}, G., {Wheeler}, J.~C., {et~al.} 2007, \apjl, 668,
  L99

\bibitem[{{Quimby} {et~al.}(2011){Quimby}, {Kulkarni}, {Kasliwal}, {Gal-Yam},
  {Arcavi}, {Sullivan}, {Nugent}, {Thomas}, {Howell}, {Nakar}, {Bildsten},
  {Theissen}, {Law}, {Dekany}, {Rahmer}, {Hale}, {Smith}, {Ofek}, {Zolkower},
  {Velur}, {Walters}, {Henning}, {Bui}, {McKenna}, {Poznanski}, {Cenko}, \&
  {Levitan}}]{quimby11}
{Quimby}, R.~M., {Kulkarni}, S.~R., {Kasliwal}, M.~M., {et~al.} 2011, \nat,
  474, 487

\bibitem[{{Quimby} {et~al.}(2013){Quimby}, {Werner}, {Oguri}, {More}, {More},
  {Tanaka}, {Nomoto}, {Moriya}, {Folatelli}, {Maeda}, \& {Bersten}}]{Quimby13}
{Quimby}, R.~M., {Werner}, M.~C., {Oguri}, M., {et~al.} 2013, \apjl, 768, L20

\bibitem[{{Rakavy} \& {Shaviv}(1967)}]{Rakavy67}
{Rakavy}, G., \& {Shaviv}, G. 1967, \apj, 148, 803

\bibitem[{{Sanders} {et~al.}(2012){Sanders}, {Soderberg}, {Levesque}, {Foley},
  {Chornock}, {Milisavljevic}, {Margutti}, {Berger}, {Drout}, {Czekala}, \&
  {Dittmann}}]{sanders12}
{Sanders}, N.~E., {Soderberg}, A.~M., {Levesque}, E.~M., {et~al.} 2012, \apj,
  758, 132

\bibitem[{{Sauer} {et~al.}(2006){Sauer}, {Mazzali}, {Deng}, {Valenti},
  {Nomoto}, \& {Filippenko}}]{Sauer06}
{Sauer}, D.~N., {Mazzali}, P.~A., {Deng}, J., {et~al.} 2006, \mnras, 369, 1939

\bibitem[{{Schlafly} \& {Finkbeiner}(2011)}]{Schlafly11}
{Schlafly}, E.~F., \& {Finkbeiner}, D.~P. 2011, \apj, 737, 103

\bibitem[{{Schlegel} {et~al.}(1998){Schlegel}, {Finkbeiner}, \&
  {Davis}}]{schlegel98}
{Schlegel}, D.~J., {Finkbeiner}, D.~P., \& {Davis}, M. 1998, \apj, 500, 525

\bibitem[{{Schulze} {et~al.}(2016){Schulze}, {Kr{\"u}hler}, {Leloudas},
  {Gorosabel}, {Mehner}, {Buchner}, {Kim}, {Ibar}, {Amor{\'{\i}}n},
  {Herrero-Illana}, {Anderson}, {Bauer}, {Christensen}, {de Pasquale}, {de
  Ugarte Postigo}, {Gallazzi}, {Hjorth}, {Morrell}, {Malesani}, {Sparre},
  {Stalder}, {Stark}, {Th{\"o}ne}, \& {Wheeler}}]{Schulze16}
{Schulze}, S., {Kr{\"u}hler}, T., {Leloudas}, G., {et~al.} 2016, ArXiv
  e-prints, arXiv:1612.05978

\bibitem[{{Shivvers} {et~al.}(2016){Shivvers}, {Zheng}, {Mauerhan}, {Kleiser},
  {Van Dyk}, {Silverman}, {Graham}, {Kelly}, {Filippenko}, \&
  {Kumar}}]{shivvers16}
{Shivvers}, I., {Zheng}, W.~K., {Mauerhan}, J., {et~al.} 2016, \mnras, 461,
  3057

\bibitem[{{Smith} {et~al.}(2016){Smith}, {Sullivan}, {D'Andrea}, {Castander},
  {Casas}, {Prajs}, {Papadopoulos}, {Nichol}, {Karpenka}, {Bernard}, {Brown},
  {Cartier}, {Cooke}, {Curtin}, {Davis}, {Finley}, {Foley}, {Gal-Yam},
  {Goldstein}, {Gonz{\'a}lez-Gait{\'a}n}, {Gupta}, {Howell}, {Inserra},
  {Kessler}, {Lidman}, {Marriner}, {Nugent}, {Pritchard}, {Sako}, {Smartt},
  {Smith}, {Spinka}, {Thomas}, {Wolf}, {Zenteno}, {Abbott}, {Benoit-L{\'e}vy},
  {Bertin}, {Brooks}, {Buckley-Geer}, {Carnero Rosell}, {Carrasco Kind},
  {Carretero}, {Crocce}, {Cunha}, {da Costa}, {Desai}, {Diehl}, {Doel},
  {Estrada}, {Evrard}, {Flaugher}, {Fosalba}, {Frieman}, {Gerdes}, {Gruen},
  {Gruendl}, {James}, {Kuehn}, {Kuropatkin}, {Lahav}, {Li}, {Marshall},
  {Martini}, {Miller}, {Miquel}, {Nord}, {Ogando}, {Plazas}, {Reil}, {Romer},
  {Roodman}, {Rykoff}, {Sanchez}, {Scarpine}, {Schubnell}, {Sevilla-Noarbe},
  {Soares-Santos}, {Sobreira}, {Suchyta}, {Swanson}, {Tarle}, {Walker},
  {Wester}, \& {DES Collaboration}}]{smith16}
{Smith}, M., {Sullivan}, M., {D'Andrea}, C.~B., {et~al.} 2016, \apjl, 818, L8

\bibitem[{{Smith} {et~al.}(2010){Smith}, {Chornock}, {Silverman}, {Filippenko},
  \& {Foley}}]{smith10}
{Smith}, N., {Chornock}, R., {Silverman}, J.~M., {Filippenko}, A.~V., \&
  {Foley}, R.~J. 2010, \apj, 709, 856

\bibitem[{{Soker}(2016)}]{Soker16}
{Soker}, N. 2016, New Astronomy, 47, 88

\bibitem[{{Sorokina} {et~al.}(2016){Sorokina}, {Blinnikov}, {Nomoto}, {Quimby},
  \& {Tolstov}}]{sorokina16}
{Sorokina}, E., {Blinnikov}, S., {Nomoto}, K., {Quimby}, R., \& {Tolstov}, A.
  2016, \apj, 829, 17

\bibitem[{{Suzuki} \& {Maeda}(2016)}]{Suzuki16}
{Suzuki}, A., \& {Maeda}, K. 2016, ArXiv e-prints, arXiv:1612.03911

\bibitem[{{Taddia} {et~al.}(2015){Taddia}, {Sollerman}, {Leloudas},
  {Stritzinger}, {Valenti}, {Galbany}, {Kessler}, {Schneider}, \&
  {Wheeler}}]{taddia15}
{Taddia}, F., {Sollerman}, J., {Leloudas}, G., {et~al.} 2015, \aap, 574, A60

\bibitem[{{Tolstov} {et~al.}(2016){Tolstov}, {Nomoto}, {Blinnikov}, {Sorokina},
  {Quimby}, \& {Baklanov}}]{Tolstov16}
{Tolstov}, A., {Nomoto}, K., {Blinnikov}, S., {et~al.} 2016, ArXiv e-prints,
  arXiv:1612.01634

\bibitem[{{Tremonti} {et~al.}(2004){Tremonti}, {Heckman}, {Kauffmann},
  {Brinchmann}, {Charlot}, {White}, {Seibert}, {Peng}, {Schlegel}, {Uomoto},
  {Fukugita}, \& {Brinkmann}}]{tremonti04}
{Tremonti}, C.~A., {Heckman}, T.~M., {Kauffmann}, G., {et~al.} 2004, \apj, 613,
  898

\bibitem[{{Vreeswijk} {et~al.}(2014){Vreeswijk}, {Savaglio}, {Gal-Yam}, {De
  Cia}, {Quimby}, {Sullivan}, {Cenko}, {Perley}, {Filippenko}, {Clubb},
  {Taddia}, {Sollerman}, {Leloudas}, {Arcavi}, {Rubin}, {Kasliwal}, {Cao},
  {Yaron}, {Tal}, {Ofek}, {Capone}, {Kutyrev}, {Toy}, {Nugent}, {Laher},
  {Surace}, \& {Kulkarni}}]{Vreeswijk14}
{Vreeswijk}, P.~M., {Savaglio}, S., {Gal-Yam}, A., {et~al.} 2014, \apj, 797, 24

\bibitem[{{Waldman}(2008)}]{Waldman08}
{Waldman}, R. 2008, \apj, 685, 1103

\bibitem[{{Wang} {et~al.}(2016){Wang}, {Liu}, {Dai}, {Wang}, \& {Wu}}]{wang16}
{Wang}, S.~Q., {Liu}, L.~D., {Dai}, Z.~G., {Wang}, L.~J., \& {Wu}, X.~F. 2016,
  \apj, 828, 87

\bibitem[{{Woosley}(2010)}]{woosley10}
{Woosley}, S.~E. 2010, \apjl, 719, L204

\bibitem[{{Woosley}(2016)}]{Woosley16}
---. 2016, ArXiv e-prints, arXiv:1608.08939

\bibitem[{{Woosley} {et~al.}(2007){Woosley}, {Blinnikov}, \&
  {Heger}}]{Woosley07}
{Woosley}, S.~E., {Blinnikov}, S., \& {Heger}, A. 2007, \nat, 450, 390

\bibitem[{{Yan} {et~al.}(2015){Yan}, {Quimby}, {Ofek}, {Gal-Yam}, {Mazzali},
  {Perley}, {Vreeswijk}, {Leloudas}, {De Cia}, {Masci}, {Cenko}, {Cao},
  {Kulkarni}, {Nugent}, {Rebbapragada}, {Wo{\'z}niak}, \& {Yaron}}]{yan15}
{Yan}, L., {Quimby}, R., {Ofek}, E., {et~al.} 2015, \apj, 814, 108

\bibitem[{{Young} {et~al.}(2010){Young}, {Smartt}, {Valenti}, {Pastorello},
  {Benetti}, {Benn}, {Bersier}, {Botticella}, {Corradi}, {Harutyunyan},
  {Hrudkova}, {Hunter}, {Mattila}, {de Mooij}, {Navasardyan}, {Snellen},
  {Tanvir}, \& {Zampieri}}]{young10}
{Young}, D.~R., {Smartt}, S.~J., {Valenti}, S., {et~al.} 2010, \aap, 512, A70+

\bibitem[{{Zahid} {et~al.}(2011){Zahid}, {Kewley}, \& {Bresolin}}]{Zahid11}
{Zahid}, H.~J., {Kewley}, L.~J., \& {Bresolin}, F. 2011, \apj, 730, 137

\end{thebibliography}

\end{document}